\documentclass[aps,amsmath,amssymb,superscriptaddress,floatfix,twocolumn]{revtex4-1}
\usepackage{amsmath}
\usepackage{changepage}
\usepackage{epsfig}
\usepackage{epstopdf}
\usepackage[colorlinks,citecolor=blue,linkcolor=blue]{hyperref}
\usepackage{lipsum}
\usepackage{booktabs}
\usepackage{multirow}
\usepackage{braket}
\usepackage{multirow}
\usepackage{makecell}
\usepackage{amsmath}
\usepackage{mathtools}

\begin{document}
\title{Adiabatic  topological passage based on coupling of giant atom with two Su-Schrieffer-Heeger chains  }
\author{Da-Wei Wang}
\affiliation{School of Integrated Circuits, Tsinghua University, Beijing 100084, China}
\author{Ling Zhou}
\affiliation{School of Physics, Dalian University of Technology, Dalian 116024, China}
\author{Yu-xi Liu}
\thanks{yuxiliu@mail.tsinghua.edu.cn}
\affiliation{School of Integrated Circuits, Tsinghua University, Beijing 100084, China}
\begin{abstract}
We study an adiabatic topological passage of  two Su-Schrieffer-Heeger (SSH) chains mediated by a giant atom.  When two finite SSH chains are in the topological phase and the frequency of the giant atom is equal to the center frequency of the SSH chains,  the system is reduced to a subsystem that describes the coupling of a giant atom  to the edge states of two SSH chains.  In this case, we can find dark states that act as adiabatic topological passages. This allows us to adiabatically transfer excitations of the giant atom to either one end of two SSH chains in a fully controllable way. In addition, we  show good robustness of the adiabatic topological passages to both giant atom frequency mismatch and the coupling disorders in two SSH chains. Our study provides a method to realize quantum information processing and fabricate quantum optical devices based on the coupling of the giant atom to topological matter.
\end{abstract}

\maketitle

\section{Introduction}
A crucial task to implement efficient quantum information processing is information coherent transfer between different nodes~\cite{Kimble2008,PhysRevLett.78.3221}.  Several quantum channels for state transfer have been proposed in different quantum systems, e.g., optical lattices~\cite{PhysRevLett.107.210405}, the systems of cavity quantum electrodynamics~\cite{PhysRevLett.78.3221,RevModPhys.87.1379,Ritter2012}, photonic and phononic  waveguides~\cite{PhysRevLett.118.133601,PhysRevX.8.041027}, superconducting quantum circuits~\cite{Axline2018,PhysRevA.98.012331}, quantum dot arrays \cite{PETROSYAN2006419}. However, due to the inevitable environmental effects of these systems,  the reliability of quantum information transfer  may be significantly reduced. Therefore, how to realize a robust channel for quantum state transfer remains an urgent and ongoing question.
Various approaches have been proposed to overcome the effects of perturbations and decoherence during coherent transfer between nodes through the quantum channel, the best-known one is the stimulated Raman adiabatic passage (STIRAP)~\cite{RevModPhys.89.015006}  in a three-levels system~\cite{PhysRevA.40.6741}, in which controllable population transfer between the initial and final states is not subject to any loss of intermediate radiative states~\cite{RevModPhys.70.1003,PhysRevA.88.022323}.

Another option is to utilize topology-protected edge states as quantum channels.  One of the most attractive properties of topological systems is that the edge states are robust to small perturbations~\cite{RevModPhys.82.3045, Wray2010,Shalaev2019,RevModPhys.91.015006}. The research along this direction can be classified into two categories. One is that the information is encoded and processed on the edge states, and  the state transfer between two edges is realized via well-known adiabatic techniques~\cite{PhysRevB.27.6083,PhysRevLett.109.106402,Longhi2019,PhysRevA.103.052409}, Landau-Zener tunnel~\cite{Longhi2019} or STIRAP~\cite{RevModPhys.89.015006,PhysRevB.99.155150,PhysRevResearch.2.033475}.  These methods are well adapted to non-diagonal disorder, but the speed of the state transfer is intrinsically limited by the adiabatic requirements, i.e., avoiding unwanted adiabatic transitions in the edge states. The Su-Schrieffer-Heeger (SSH) model and its various extensions have been investigated in  this scenarios \cite{PhysRevLett.129.215901,PhysRevA.107.053701,PhysRevA.103.052409,PhysRevA.98.012331,PhysRevA.106.052411,PhysRevA.102.022404,PhysRevA.103.023504}. Beyond the fast and robust quantum state transfer based on topological edge state channels, a large amount of work has been devoted to study topological quantum devices for robust information processing, e.g.,  topological beam splitters, routers and other combinations~\cite{PhysRevApplied.18.054037,PhysRevA.107.062214,PhysRevResearch.3.023037,PhysRevB.103.085129}.
 Another topology-based channel is to couple the quantum system with topological systems, in which their couplings are manipulated to realize topology-preserving quantum information processing~\cite{PhysRevA.106.052411,PhysRevB.109.205421,PhysRevLett.124.023603,PhysRevResearch.2.012076,Dlaska_2017,Yao2013,PhysRevA.106.033710,PhysRevA.106.052411,PhysRevA.108.023703,PhysRevA.109.033708}. For example, a topologically protected channel has been proposed for quantum state transfer between remote quantum nodes, in which the quantum state transfer is mediated by edge modes of a chiral spin liquid, and this method is intrinsically robust to realistic defects associated with disorder and decoherence \cite{Yao2013}.

Recently,  giant atoms have been studied~\cite{Andersson2019,Kannan2020,PhysRevX.13.021039,PhysRevA.103.023710} for coupling them to can be to the waveguide via several different coupling points. This results in many interesting phenomena, such as frequency-dependent decay rates and Lamb shifts~\cite{PhysRevA.90.013837,PhysRevLett.123.233602}, decoherence-free interaction ~\cite{PhysRevLett.120.140404,PhysRevResearch.2.043184,PhysRevA.107.023705}, chiral phenomena~\cite{PhysRevA.105.023712,PhysRevLett.126.043602}, oscillating bound states~\cite{PhysRevA.102.033706,PhysRevResearch.2.043014}, and non-Markovian effects~\cite{PhysRevLett.133.063603,PhysRevA.106.033522,Andersson2019}.  These new phenomena arise from the interference and time-delay effects of the electromagnetic fields propagating between coupling points. In addition, the giant atom can also be coupled to several waveguides for implementing controllable transmission of photons between different waveguides~\cite{Chen2022}.
Here, we consider a giant atom, which is coupled to two finite SSH chains. We propose to realize directional quantum state transfer by constructing the adiabatic topological passage.  When the two finite SSH chains are in topological phase, the whole system is reduced to a subsystem that describes the coupling of a giant atom  to the edge states of two SSH chains.
Then, the subsystem can be equivalently described by a five-state model consisting of the edge states of the two SSH chains as well as the giant atom. In this case, we find that there is a dark state in the subsystem that can act as an adiabatic topological passage, allowing us to adiabatically transfer the excited state of the giant atom to either end of the two SSH chains in a fully controllable way. Moreover, we discuss the effects of the giant atom frequency mismatch and non-diagonal disorder in SSH chains, and the results show that our proposal  exhibits good robustness to both of these imperfections.

The paper is organized as follows: In Sec.~\ref{I}, we describe a theoretical model and derive the effective Hamiltonian for realizing the coupling between giant atom and edge states of the  two SSH chains.
In Sec.~\ref{II},  We construct adiabatic topological passages and apply them to quantum information transfer.  In Sec.~\ref{III}, we discuss the imperfection effect on the adiabatic topological passages. In Sec.~\ref{IV}, we further discuss our result and experimental feasibility. Finally, we conclude our results in Sec.~\ref{V}.

\begin{figure}[t]
  \centering
  \includegraphics[width=8cm]{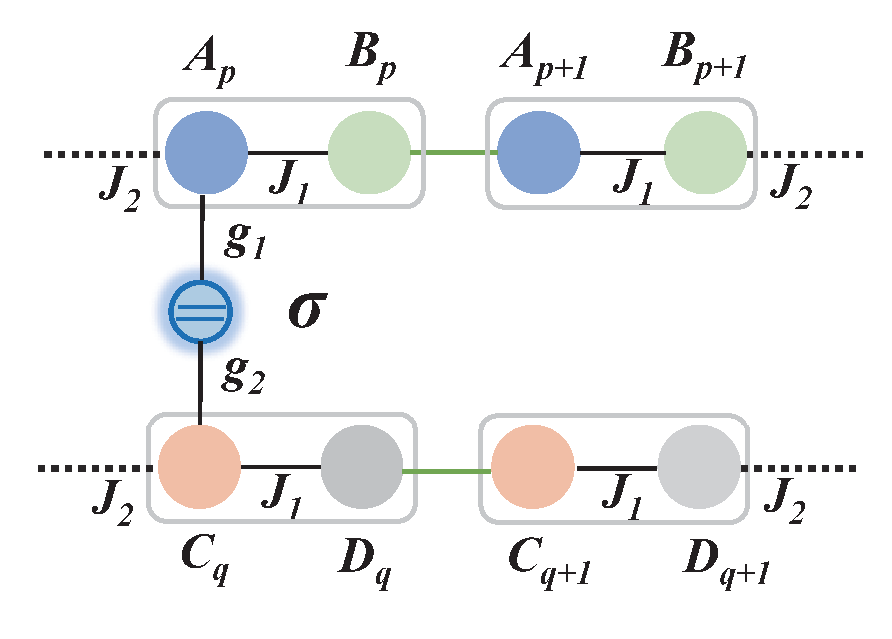}
  \caption{Schematic diagram for the coupling of a giant atom to two finite SSH chains at sublattice $A_p$ and $C_q$ with coupling strength $g_1$ and $g_2$, respectively. Alternatively, the giant atom can also be coupled to the  sublattice $B_p$ or $D_q$  (not shown here).}
  \label{model}
\end{figure}

\section{atom coupling with the finite SSH chain}\label{I}

As schematically shown in Fig.~\ref{model}, we consider the system that a two-level giant atom is coupled to two finite SSH chains denoted by SSH-$1$ and SSH-$2$, respectively. The giant atom is coupled to the SSH-$1$ chain  via either the sublattice $A_p$ or the sublattice $B_p$ of the $p$th cell with the coupling strength $g_1$. However, the coupling to the SSH-$2$ chain is realized via either sublattice $C_p$ or the sublattice $D_p$  of  the $q$th cell with the coupling strength $g_2$. The total Hamiltonian can be written as
\begin{equation}
  \begin{aligned}
    H=\omega_e \sigma^\dag\sigma^{-}+H^1_{\mathrm{SSH}}+H^2_{\mathrm{SSH}}+[(g_1x_{p}^{\dagger}+g_2y_{q}^{\dagger})\sigma+\rm{h.c.}],
  \end{aligned}\label{ee1}
  \end{equation}
where $\sigma^\dag=|e\rangle\langle g|$ and $\sigma^{-}=|g\rangle\langle e|$  are the raising  and lowering operators of the giant atom with the frequency $\omega_e$ from the ground $|g\rangle$ to the excited $|e\rangle$ state. The operator $x^\dagger_p=a^\dag_p$ or $b^\dag_p$  ($y^\dagger_q=c^\dag_q$ or $ d^\dag_q$) represents the  creation operator in the sublattice  $A_p$ or  $B_p$ ($C_q$ or $D_q$) with the possible coupling of the giant atom to the SSH-$1$ (SSH-$2$) chain. There are four possible coupling scenarios $(a^\dag_p,c^\dag_q)$,  $(a^\dag_p,d^\dag_q)$, $(a^\dag_p,c^\dag_q)$ and $(b^\dag_p,d^\dag_q)$. For example, the scenario $(a^\dag_p,c^\dag_q)$ denotes that the giant atom is coupled to the sublattice $A_p$ of the $p$th cell of the SSH-$1$ chain and  the sublattice $C_p$ of the $q$th cell of the SSH-$2$ chain.
The Hamiltonians $H^1_{\mathrm{SSH}}$ and $H^2_{\mathrm{SSH}}$ in Eq.~(\ref{ee1}) corresponding to two SSH chains   are
\begin{subequations}
  \begin{align}
    \label{eq2a}
    H^1_{\mathrm{SSH}}&=\sum_{i=1}^N\omega_o(a_{i}^{\dagger}a_{i}+b_{i}^{\dagger}b_{i}) +(J_1a_{i}^{\dagger}b_i+ J_2a_{i+1}^{\dagger}b_i+\rm{h.c.}),\\ \label{eq2b}
    H^2_{\mathrm{SSH}}&=\sum_{i=1}^N\omega_o(c_{i}^{\dagger}c_{i}+d_{i}^{\dagger}d_{i}) +(J_1c_{i}^{\dagger}d_i+ J_2c_{i+1}^{\dagger}d_i+\rm{h.c.}).
  \end{align}
\end{subequations}
Here, we assume that all sublattices have the same frequency $\omega_{o}$ such that the quantized Zak phase in the SSH model is still valid~\cite{PhysRevB.84.195452,PhysRevLett.62.2747}. We define $\omega_o$ as the central the frequency  of  the SSH chains and consider it  as the reference in the following discussions. The intra-cell sublattice jump and inter-cell sublattice hopping strengths in the SSH-1(SSH-2) chain are $J_1=J(1+\cos\theta)$ and $J_2=J(1-\cos\theta)$  with $\theta \in [0,2\pi]$, respectively. For the finite-sized SSH chain, when the hopping strength of the SSH model satisfies $J_1>J_2$, the SSH model is in the trivial phase, in this case, there exists only one finite energy gap in the energy spectrum; when the hopping strength satisfies $J_1<J_2$, the SSH model is in the topological phase in which there exists two degenerate states in the energy gap,  arising from the hybridization of the left and right edge states due to finite-size effects~\cite{Asboth2015}. These two  band-gap states are topology-protected, and thus utilizing topology-protected states to transfer information may be robust to a certain range of disorders.

In the single-excited space, we label $|A_i\rangle=a_i^\dag|G\rangle$ and $|B_i\rangle=b_i^\dag|G\rangle$  ($|C_i\rangle=c_i^\dag|G\rangle$ and $|D_i\rangle=d_i^\dag|G\rangle$), where $|G\rangle$ is the ground state of the SSH-$1$ (SSH-$2$) chain.
We can diagonalize the Hamiltonian in Eqs.~(\ref{eq2a}) and (\ref{eq2b}) as $H^1_{\mathrm{SSH}}=\sum_{j=1}^{2N}E_j|\Psi_j\rangle\langle\Psi_j|$, where $E_j$ and $|\Psi_j\rangle$ are the eigenenergies and corresponding eigenvectors with $E_{2N}>E_{(2N-1)}>\cdots>E_{1}$;  $H^2_{\mathrm{SSH}}=\sum_{j=1}^{2N}\varepsilon_j|\Phi_j\rangle\langle\Phi_j|$ with the eigenenergies  $\varepsilon_j$ and the  eigenvectors $|\Phi_j\rangle$ with $\varepsilon_{2N}>\varepsilon_{(2N-1)}>\cdots>\varepsilon_{1}$. It can be verified that $E_{N}\approx E_{N+1}\approx\varepsilon_{N}\approx \varepsilon_{N+1}\approx 0$ in the topological phase for the finite length of the chains when $\omega_o$ is taken as the reference energy.
For scenario $(a^\dag_p,c^\dag_q)$, we can rewrite the Hamiltonian~(\ref{ee1}) as
\begin{equation}
  \begin{aligned}
   H&=\Delta \sigma^\dag\sigma^-+\sum\limits_{j=1}^{2N}\left [E_j|\Psi_j\rangle\langle\Psi_j|+\varepsilon_j|\Phi_j\rangle\langle\Phi_j| \right.\\ &\left.\quad+(g_1\eta_{2p-1,j}\Psi_j^\dag+g_2\zeta_{2q-1,j}\Phi_j^\dag)\sigma^-+\rm{h.c.}\right], \label{e5}
\end{aligned}
\end{equation}
with  $\Delta=\omega_e-\omega_o$,  $\Psi_j^\dag=|\Psi_j\rangle\langle G|$ and $\Phi_j^\dag=|\Phi_j\rangle\langle G|$. $g_{1}\eta_{2p-1,j}=g_{1}\langle\Psi_j|A_p\rangle$ ($g_2\zeta_{2q-1,j}=g_2\langle\Phi_j|C_q\rangle$) is the coupling strength between the giant atom and the SSH-$1$ (SSH-$2$) chain in the diagonalized basis. Hereafter, $g_1\eta_{2p-1,j}$ (or $g_1\eta_{2p,j}$) is the coupling strength when  the giant atom is coupled to the sublattice $A$ (or $B$) of the $p$th unit cell in the SSH-$1$, and $g_2\eta_{2q-1,j}$ (or $g_2\eta_{2q,j}$) is the coupling strength when the giant atom is coupled to the sublattice $C$ (or $D$) of the $q$th unit cell in the SSH-$2$.

\begin{figure}[t]
  \centering
  \includegraphics[width=8cm]{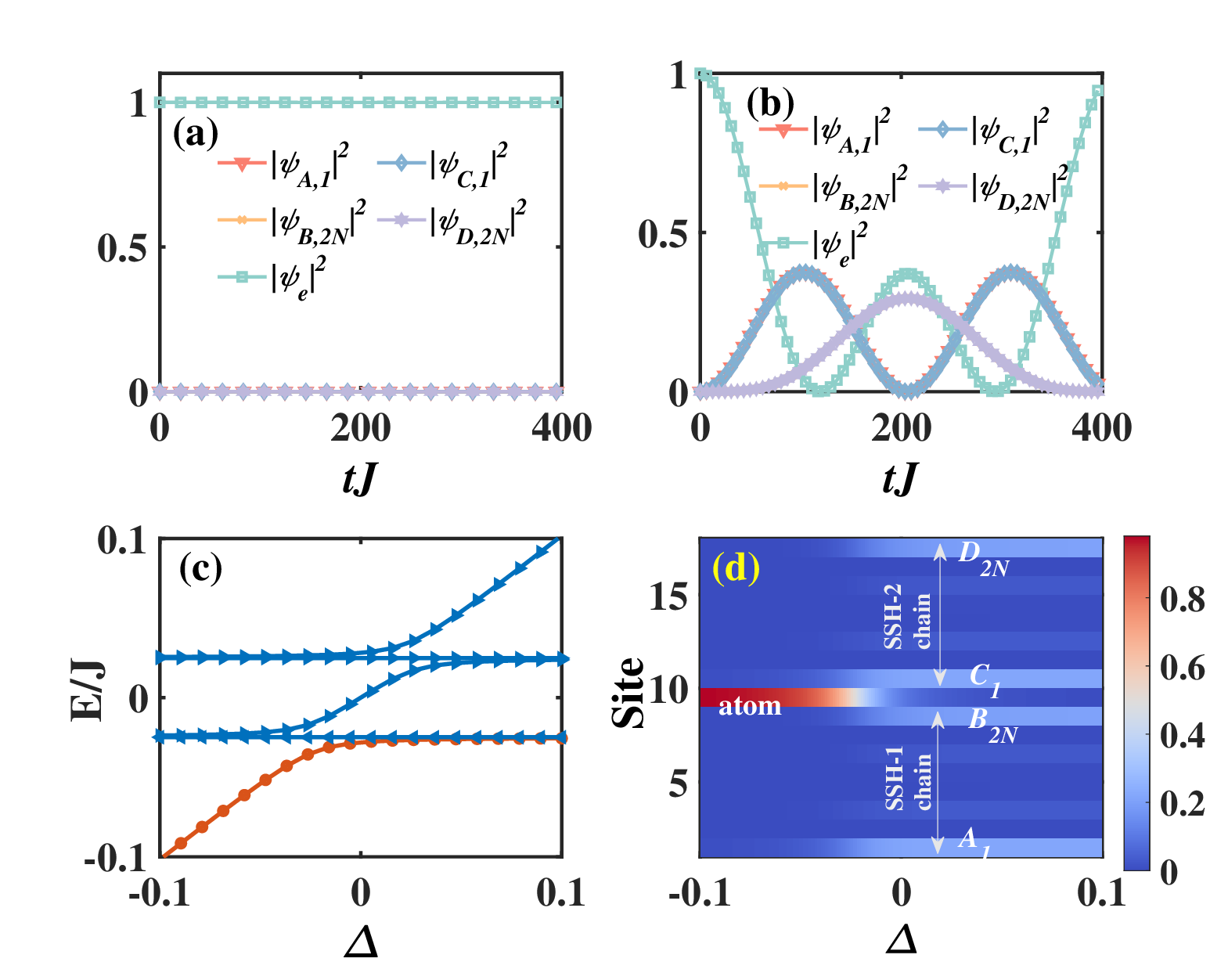}
  \caption{ The  probability evolution of  the atomic excited state $|\psi_e|^2$, sublattice $A_1$ ($|\psi_{A,1}|^2$), sublattice $B_{2N}$ ($|\psi_{B,2N}|^2$),  sublattice $C_1$ ($|\psi_{C,1}|^2$) and  sublattice $D_{2N}$ ($|\psi_{D,2N}|^2$) with $\Delta=0$ for (a) $\theta=0.4\pi$ (trivial phase) and for (b) $\theta=0.7\pi$ (topological phase). (c) The partial  energy spectrum of the system varies with $\Delta$ for the two SSH chains in the topological phase.
  (d) The distribution of the probability of hybridized mode that the circle  marker in panel (c) varies with $\Delta$. The  parameters are $p=q=1$,  $N=4$,  $g_1=g_2=0.01J$ and $J=1$.}
  \label{detuning_energy_evolution}
 \end{figure}

In the interaction picture by performing the unitary transformation
\begin{equation}
U=\exp\left[-i\left(\sum\limits_{j=1}^{2N}E_j\Psi_j^\dag\Psi_j+\varepsilon _j\Phi_j^\dag\Phi_j+\Delta \sigma^\dag\sigma^-\right)t\right],
\end{equation}
the Hamiltonian (\ref{e5}) can be written as
\begin{equation}
  \begin{aligned}
   H_{I}&=\sum\limits_{j=1}^{2N}\left [g_1\zeta_{2p-1,j}e^{i(E_j-\Delta)t}\Psi_j^\dag\sigma^- \right.\\
   &\left. \quad+g_2\eta_{2q-1,j}e^{i(\varepsilon_j-\Delta)t}\Phi_j^\dag\sigma^-+\rm{h.c.}\right] . \label{e6}
\end{aligned}
\end{equation}
 Below, we consider that the giant atom resonantly interacts  with the SSH-$1$ and  SSH-$2$ chains with $\Delta = 0$ and the condition  $\{g_1/J,g_2/J\} \ll 1$.  For the SSH-$1$ and SSH-$2$ chains in the trivial phase, the giant atom nearly decouples from the chain due to $|E_j|\gg g_1$ and  $|\varepsilon _j|\gg g_2$. In Fig.~\ref{detuning_energy_evolution}(a), we plot the  dynamical evolution of the system when the system is initially in the excited state $|e\rangle$ of the giant atom and the vacuum state of the two SSH chains with $\theta=0.4\pi$. It can be found that the excitation remains in the atom and is not exchanged with two SSH chains. For the SSH-$1$ and  SSH-$2$ chains in the topological phase, due to the presence of four band-gap states corresponding to nearly zero energy $\{E_{N}, E_{N+1},\varepsilon _{N}, \varepsilon _{N+1}\} \approx 0$, which are near resonant with the atomic frequency for $\Delta=0$, then the coupling between other eigenstates and the giant atom can be neglected due to $|E_{j}|\gg  g\eta_{2p-1,j}>0$ and $|\varepsilon_{j}|\gg  g\zeta_{2q-1,j}$ for $j\neq N, N+1$. Under this case,  the effective interaction Hamiltonian of the giant atom and the SSH chains in the Schr\"{o}dinger's picture can be written as
 \begin{equation}
   \begin{aligned}
    H&=E_N\Psi_{N}^\dag\Psi_{N}+E_{N+1}\Psi_{N+1}^\dag\Psi_{N+1}\\ &\quad +\varepsilon _N\Phi_{N}^\dag\Phi_{N}+\varepsilon _{N+1}\Phi_{N+1}^\dag\Phi_{N+1}\\ &\quad +[g_1(\eta_{2p-1,N}\Psi_N^\dag+\eta_{2p-1,N+1}\Psi_{N+1}^\dag)\sigma^-\\ &\quad g_2(\zeta_{2q-1,N}\Phi_N^\dag+\zeta_{2q-1,N+1}\Phi_{N+1}^\dag)\sigma^-+\rm{h.c.}]. \label{e16}
 \end{aligned}
 \end{equation}

 That is, in the single-excitation space, when we consider that the giant atom is resonant with the SSH chain, the above Hamiltonian  only causes the following subspace transitions $\{|\psi_N,g\rangle,|\phi_N,g\rangle \}\leftrightarrow |vac, e\rangle$ $\leftrightarrow \{|\psi_{N+1},g\rangle,|\phi_{N+1},g\rangle \}$. We also plot the dynamical evolution of the system in Fig.~\ref{detuning_energy_evolution}(b) with the same parameters as in Fig.~\ref{detuning_energy_evolution}(a) except $\theta=0.7\pi$. We can find that the  excited state of the giant atom can be transferred to  the end lattices of the two SSH chains.  We have also plotted the energy spectrum of the system corresponding to the Hamiltonian in Eq.~(\ref{ee1})  with the variation of the detuning $\Delta$  in Fig.~\ref{detuning_energy_evolution}(c), where we show only the range of $E\in[-0.1J, 0.1J]$  to see the details clearly.
 In this energy range, there are only five energy levels. It can be seen that the energy levels show an anti-crossing property.  That is,  there are coherent couplings~\cite{PhysRevA.98.023841,PhysRevLett.114.227201} between the giant atom and two SSH chains, and the giant atom is coupled to the band-gap states in the two SSH chains. In Fig.~\ref{detuning_energy_evolution}(d),  the probability distribution of  the hybridization mode on the sites is plotted with the variation of $\Delta$. One can see that  the contribution of the hybridization mode gradually varies from the giant atom to the two SSH chains with the increase of $\Delta$. It also shows that   an effective transition between the giant atom and the SSH chain exists.

For an SSH chain in the topological phase, there are two band-gap states,  which are the hybridization of the left and right edge states due to finite size effects~\cite{Asboth2015} of the chain. The energies of  two band-gap states can be calculated as $E_{N}=-E_{N+1}=N_L^2(-1)^{N+1}J_1(J_1/J_2)^{N-1}$ ( $\varepsilon_N=-\varepsilon_{N+1}=N_L^2(-1)^{N+1}J_1(J_1/J_2)^{N-1}$) and the corresponding wavefunction can be written as~\cite{Asboth2015}
 \begin{equation}
 \begin{aligned}
    &|\Psi_{N}\rangle =\frac{1}{\sqrt{2}}(|\psi_L\rangle+|\psi_R\rangle),|\Psi_{N+1}\rangle =\frac{1}{\sqrt{2}}(|\psi_L\rangle-|\psi_R\rangle),\\
    &|\varPhi_{N}\rangle =\frac{1}{\sqrt{2}}(|\varphi_L\rangle+|\varphi_R\rangle),|\varPhi_{N+1}\rangle =\frac{1}{\sqrt{2}}(|\varphi_L\rangle-|\varphi_R\rangle),
 \end{aligned}\label{edge}
 \end{equation}
where $|\psi_L\rangle=N_L\sum_{i=1}^N \alpha_i^a|A_i\rangle$ ($|\varphi_L\rangle=N_L\sum_{i=1}^N \alpha_i^c|C_i\rangle$)  with  $\alpha_i^{a/c}=(-J_1/J_2)^{i-1}$ and  $|\psi_R\rangle=N_R\sum_{i=1}^N \alpha_i^b|B_i\rangle$ ($|\varphi_R\rangle=N_R\sum_{i=1}^N \alpha_i^d|D_i\rangle$) with  $\alpha_i^{b/d}=(-J_1/J_2)^{N-i}$ are the left and right edge states of the SSH-$1$ (SSH-$2$) chain.  $N_L=N_R=[1-(J_1/J_2)^2]^{1/2}[1-(J_1/J_2)^{2N}]^{-1/2}$ are the normalization constants.
Using Eq.~(\ref{edge}), we can calculate the coupling strength of the giant atom to the SSH chains in the diagonalized basis as
$g_1\eta_{2p-1,N}=-g_1\eta_{2p-1,N+1}=g_1c_p^a/\sqrt{2}$ and  $g_2\zeta_{2q-1,N}=-g_2\zeta_{2q-1,N+1}=g_2c_q^c/\sqrt{2}$.

 In the same way, we can obtain the effective coupling strengths of the giant atom to the sublattices $B_p$ and $D_q$ respectively as $g_1\eta_{2p,N}=-g_1\eta_{2p,N+1}=g_1c_p^b/\sqrt{2}$ and  $g_2\zeta_{2q,N}=-g_2\zeta_{2q,N+1}=g_2c_q^d/\sqrt{2}$.  We can see that these couplings depend on the coupling positions of the giant atom to the chains. For example, for
$\eta_{2p-1,N}=(-J_1/J_2)^{p-1}\sqrt{2}$, it decreases as $p$ increases when $J_1<J_2$, while  $\eta_{2p,N}=(-J_1/J_2)^{N-p}/\sqrt{2}$ increases with the increase of $p$. Then, the effective coupling strength of the giant atom to the SSH chains can be adjusted by changing the coupling position.

 \section{Directional and controllable excitation transfer through the adiabatic process}\label{II}
\begin{figure}[h]
  \centering
  \includegraphics[width=8cm]{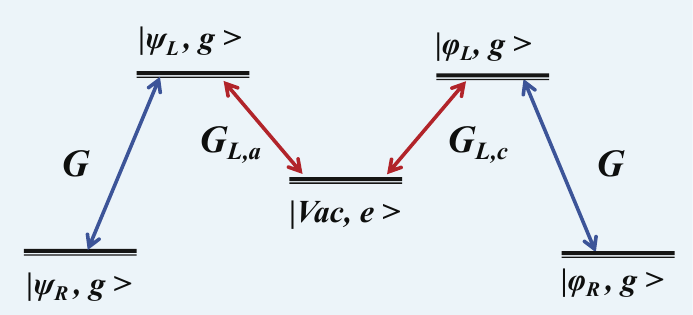}
  \caption{Schematic diagram of the transition of the subspace states $\{|vac, e\rangle$, $| \psi_{L}, g\rangle$, $| \psi_{R}, g\rangle$,$| \varphi_{L}, g\rangle$, $| \varphi_{R}, g\rangle\}$.
    }\label{five}
 \end{figure}

Now, substituting Eq.~(\ref{edge}) into Eq.~(\ref{e6}), the effective interaction Hamiltonian  between two the SSH chains in the topological phase and giant atom can be rewritten as
\begin{equation}
  \begin{aligned}
    H_{sub}&=G|\psi_L,g\rangle\langle \psi_R,g|+G|\varphi_L,g\rangle\langle \varphi_R,g|\\ &\quad+G_{L,a}|\psi_L,g\rangle\langle vac,e|+G_{L,c}|\varphi_L,g\rangle\langle vac,e|+\rm{h.c.},
   \end{aligned} \label{eq7}
\end{equation}
in the subspace $\{|vac, e\rangle$, $| \psi_{L}, g\rangle$, $| \psi_{R}, g\rangle$,$| \varphi_{L}, g\rangle$, $| \varphi_{R}, g\rangle\}$, where $G=(-1)^{N+1}N_L^2J_1(J_1/J_2)^{N-1}$ is the hybridization coupling strength between the left and right edge states due to finite size effects. $G_{L,a}=g_1N_L(-J_1/J_2)^{p-1}$ and $G_{L,c}=g_2N_L(-J_1/J_2)^{q-1}$ are the effective coupling strengths between the giant atom and two SSH chains and decay with the power exponent of $p$ and $q$, respectively. The energy state transitions described in Eq.~(\ref{eq7}) are shown schematically in Fig. ~(\ref{five}).

We can find that the giant atom is equivalently coupled to the left edge states $|\psi_L,g\rangle$ and $|\varphi_L,g\rangle$ when the giant atom is coupled to the sublattices $A_p$ and $C_q$.  Thus, we can conclude that when the giant atom is resonantly coupled to the two SSH chains in the topological phase, the full system containing the giant atom and two SSH chains can be reduced to a five-state model. And,  the coupling of   the giant atom to edge states provides us with a way to manipulate edge states through  an atom. Rewriting  Eq.~(\ref{eq7}) in the matrix form,  we can obtain
 \renewcommand\arraystretch{1.5}
 \begin{equation}
H_{\rm sub} =\left [ \begin{matrix}
 0&  G_{L,a}& 0&  G_{L,c}& 0 \\
 G_{L,a}&  0& G&  0& 0 \\
 0&  G& 0&  0& 0 \\
 G_{L,c}&  0& 0&  0& G \\
0&  0& 0&  G& 0 \\
  \end{matrix} \right ].\label{H_{eff}}
 \end{equation}

By using the eigenvalue equation corresponding to the Hamiltonian $H_{\rm sub}$ in Eq.~(\ref{H_{eff}})
 \begin{equation}
  \lambda(\lambda^2-G^2)[\lambda^2-(G^2+G_{L,a}^2+G_{L,c}^2)]=0, \label{e8}
 \end{equation}
 we can obtain $5$ eigenvalues $\lambda_0=0$, $\lambda_{\pm 1}=\pm G$  and $\lambda_{\pm 2}=\pm\sqrt{G^2+G^2_{L,a}+G^2_{L,c}}$.  We can also obtain the eigenstate corresponding to the eigenvalue $\lambda_0=0$ as
\begin{equation}
  |\Psi_0\rangle=\{\cos\chi ;0;\sin\chi\cos\phi ;0;\sin\chi\sin\phi \}, \label{zero}
 \end{equation}
where    $\tan\chi=\sqrt{G^2_{L,a}+G^2_{L,c}}/G$ and $\tan\phi=G_{L,c}/G_{L,a}$ are the mixing angles.
It can be seen that the zero-energy eigenstate $|\Psi_0\rangle$ is a superposition of the atomic excited states (with probability $\cos^2\chi$), the right edge state of the SSH-$1$ chain (with probability $\sin^2\chi\cos^2\phi$) and the right edge state of the SSH-$2$ chain (with probability $\sin^2\chi\sin^2\phi $).  Actually, the zero-energy eigenstate $|\Psi_0\rangle$ is the so-called dark state or coherent population trapping state,  and the adiabatic evolution of the zero energy state is analogous to the stimulated Raman adiabatic passage process~\cite{RevModPhys.89.015006}. However, two topological states are involved, thus we here call this adiabatic process as the adiabatic topological passage.

\begin{figure}[t]
  \centering
  \includegraphics[width=9cm]{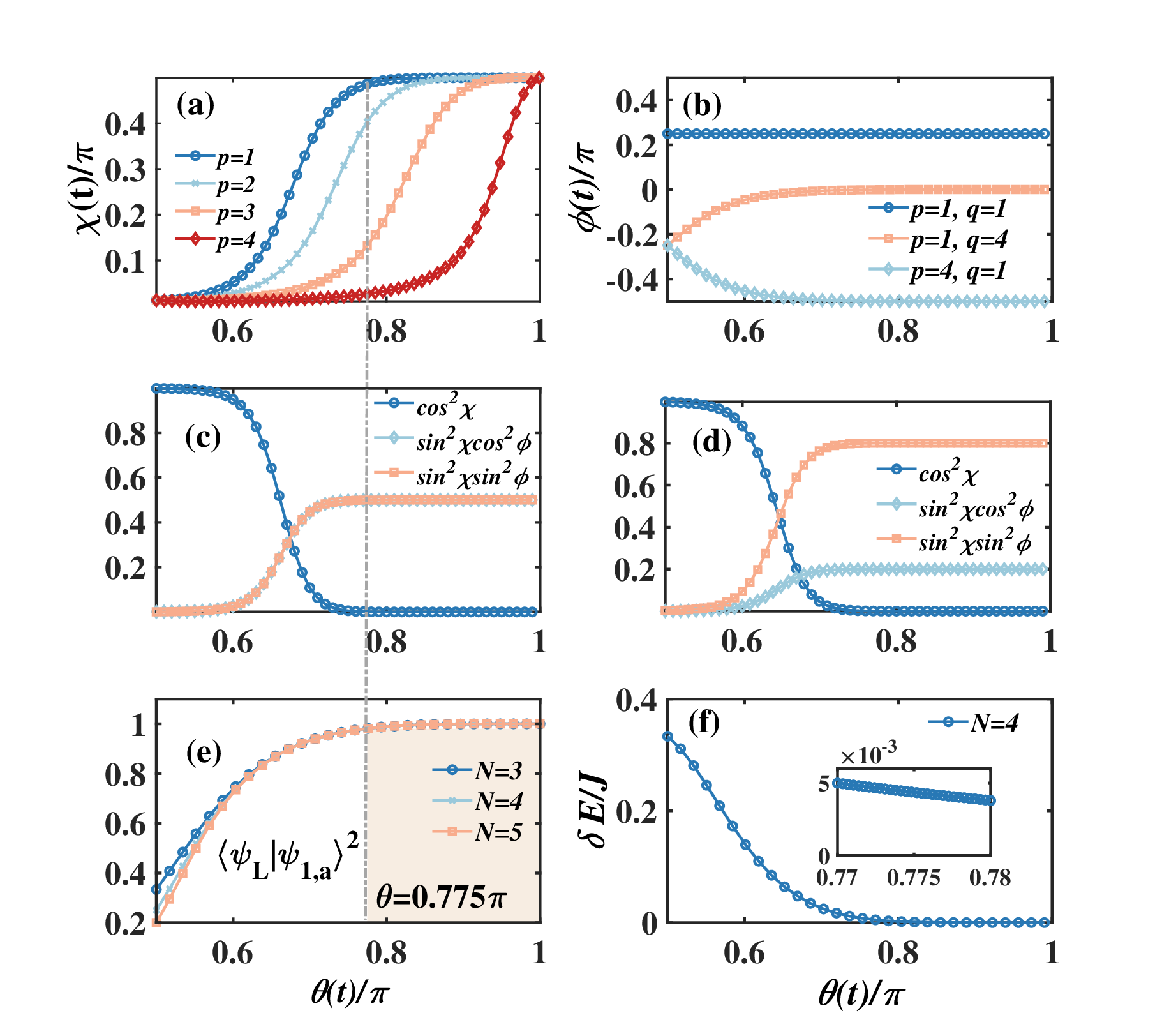}
  \caption{ (a) Mixing angle $\chi(t)$  as a function of $\theta(t)$ for $q=4$ and different $p$. (b) Mixing angle $\phi(t)$  vary with  $\theta$ for $p=q=1$ (circle), $p=1, q=4$ (square), and $p=4, q=1$ (diamond). The parameter $g_1=g_2=0.01J$ for panels (a-b).  The probability distribution of the zero energy state in giant atomic excited state (circle), right edge state of SSH-1 chain (diamond),  and right edge state of SSH-2 chain (square) versus $\theta(t)$ with  $p=q=1$ for $g_2=0.01J$  (c)  and $ g_2=0.02J$  (d), respectively; for both (c) and (d) $g_1=0.01J$.   For panel (a-d) $N=4$.  (e) The distribution of the probability that the right edge state of SSH-1 chain is in the rightmost sublattice $A_{1}$ for different $N$.  (f) The energy level difference $\delta E$ varies with $\theta(t)$.  }\label{energy_p}
 \end{figure}

 Let us assume that $\theta(t)=\Omega t$ is time-dependent with rate $\Omega$, then the couplings become  $J_1=J(1+\cos\Omega t)$ and $J_2=J(1-\cos\Omega t)$.  The time-dependent mixing angles can be obtain   $\chi(t)=\arctan(\sqrt{G^2_{L,a}(t)+G^2_{L,c}(t)}/G(t))$ and $\phi=\arctan (G_{L,c}(t)/G_{L,a}(t))$. In Fig.~\ref{energy_p}(a), we plot the variation of $\chi(t)$ with $\theta(t)$ for $q=4$ and different $p$. It can be found that $\chi(t)$ changes from $\chi(t)=0$ to $\chi(t)=\pi/2$ as $\theta(t)$ increases, implying that the atomic excited state is transferred to the right edge states of the SSH-$1$ and SSH-$2$ chains. Moreover, the reason that the transfer time depends on the atomic coupling position $p$ is that when we choose $q=4$, $G^2_{L,c}(t)$ will be close to $0$, and then as $p$ is increased, $\chi(t)$ becomes small for a fixed $\theta(t)$, and the corresponding state transfer time also increases. As for $q=1$, one can show that the transfer time is independent of $p$,  which we don't show here.   For $\tan\varphi=G_{L,a}/G_{L,c}$, it can be simplified as $\tan\phi=g_2/g_1(-J_1/J_2)^{q-p}$.   Obviously, $\phi$ is a constant for $g_1=g_2$ and $p=q$. For $g_2=g_1$, we find that $\phi$ eventually approaches 0 with the increase of the evolution time $t$ when $p<q$ and $-\pi/2$ when $p>q$ as shown in Fig.~\ref{energy_p}(b).  Because the distribution of probabilities of the two right edge states is proportional to $\tan^2\phi$,  thus we can adjust the coupling positions between the giant atom and the SSH chains to change such distribution. For example, if $p=1, q=4$, then $\phi(t)=0$ at the end of transfer time. This means that the excitation of the giant atom is only  transferred to the right edge state of the SSH-$1$ chain and the SSH-$2$ chain is always in the vacuum state.

 More particularly,  if $p=q$, then $\tan^2\phi=g_2^2/g_1^2$, the transfer of atomic excitations to the right edge states of the two SSH chains depends only on the ratio between the atom-chain coupling strengths. In Figs.~\ref{energy_p}(c) and (d), we display the  probability distribution of the zero energy state in giant atomic excited state (circle), right edge state of SSH-$1$ chain (diamond), and right edge state of SSH-$2$ chain (square) versus $\theta(t)$  for  $g_2=0.01J$ and $g_2=0.02J$, respectively.
We can see that the probability of the right state of two SSH chains is proportional to the coupling modulus.
 Furthermore, with the variation of $\theta(t)$ from $\pi/2$ to $\pi$, the probability of the right edge state $|\psi_R\rangle$ for the SSH-$1$ chain is mainly distributed on the right sublattice $B_{2i}$, and is eventually distributed on the rightmost  sublattice $B_{2N}$ when $\theta(t)$ is up  to $\theta(t)=0.775\pi$ as shown in Fig.~\ref{energy_p}(e). In addition, the transfer time from the right edge state to $B_{2N}$ is independent of  $N$.
 Combining with Figs.~\ref{energy_p}(c-e), we can conclude that  the  giant atomic excitations can be transferred to the rightmost sublattices $B_{2N}$ or $D_{2N}$ of the two SSH chains by adjusting $\theta(t)$.

 \begin{figure}[t]
  \centering
  \includegraphics[width=9cm]{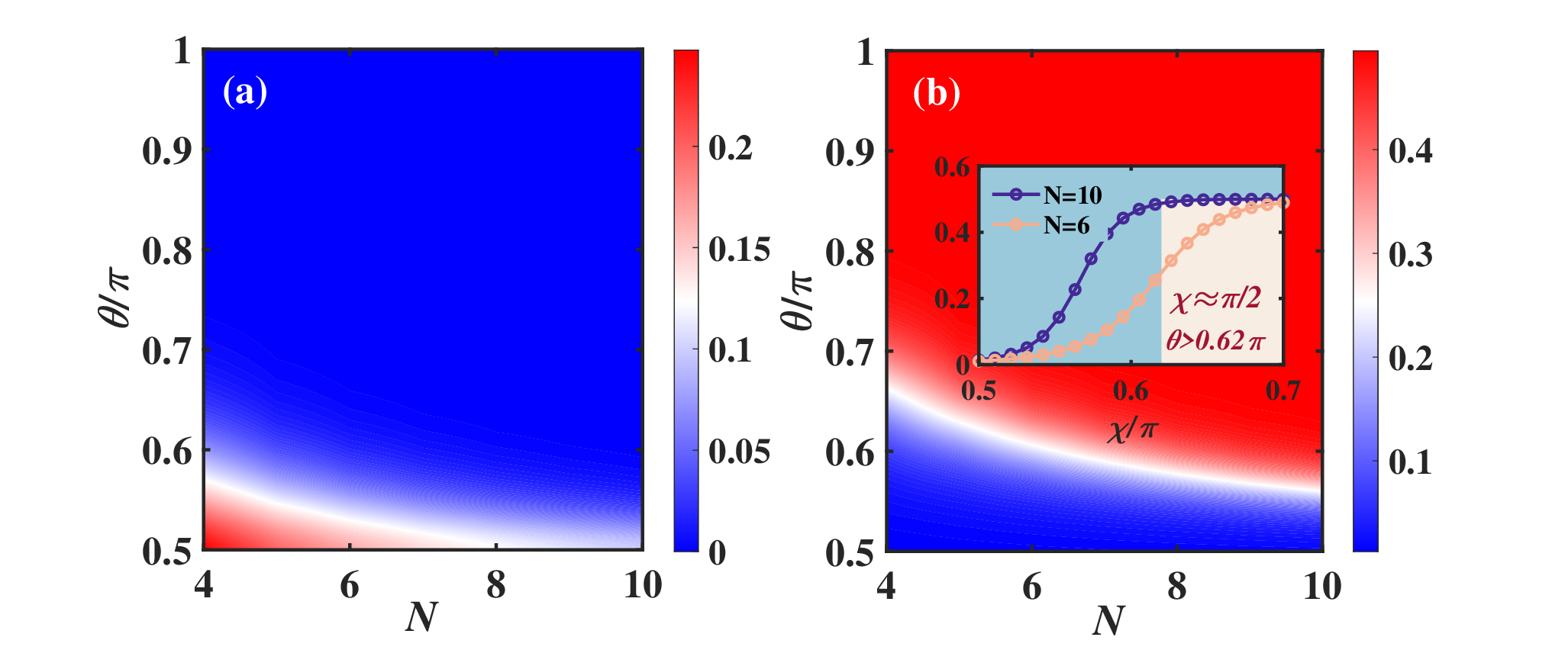}
  \caption{
  $|G|$ in (a) and $\chi/\pi$ in (b)  as a function of $N$ and $\theta$.  The insert of the panel (b) shows $\chi$ varying with $\theta$ for $N=6, 10$.  The parameters are $g_1=g_2=0.01J$, $\Delta=0$ and $J=1$.}\label{Re_A1}
  \end{figure}

 It is essential to note that the entire process needs to be adiabatic. In order to assure the adiabatic evolution, we introduce the adiabaticity parameter \cite{PhysRevA.88.022323}
 \begin{equation}\label{La}
  \varLambda (t)=\left|\frac{\langle\dot{\Psi}_G|\Psi_0\rangle}{\lambda_1(t)-\lambda_0(t)}\right|,
 \end{equation}
where $|\Psi_G\rangle$ is the eigenstate corresponding to eigenvalue $\lambda_{+1}$. For adiabatic evolution, the key requirement is that $\varLambda (t)\ll1$ throughout the evolution, which suppresses the transition from $|\Psi_0\rangle$ to $|\Psi_G\rangle$ during the evolution. In other words, this condition is equivalent to the fact that the zero-energy state $|\Psi_0\rangle$ and its nearest-energy state $|\Psi_G\rangle$ are not degenerate during the time evolution.  Therefore, we only need to make sure that there is no level crossing after the excitation is transferred from the atom to the right edge states.

In Fig.~\ref{energy_p}(f), we plot the difference between the zero-energy  and the nearest energy labeled as  $\delta E=\lambda_{+1(t)}-\lambda_0(t)=G$ for different $N$. We can see that for  $N=4$, $\delta E>10^{-3}$ when $\theta(t)=0.775\pi$ shown in the inset of Fig.~\ref{energy_p}(f), which  means that the energy levels are not degenerate and satisfy the adiabatic evolution.
We also discuss whether large $N$ allows us to utilize adiabatic evolution. Combined with the above analysis, for a given number $N$, the adiabatic evolution is valid as long as $|G|\neq0$ can be guaranteed  during the slow change of $\theta$.
In Fig.~\ref{Re_A1}(a), we display the energy difference $\delta E=|G|$  as a function of $N$ and $\theta$.  One can see that $|G|$  decreases as $N$ and $\theta$ increase.  Meanwhile,  $\chi$ as a function of $N$ and $\theta$ is shown in Fig.~\ref{Re_A1}(b).  It can be found that $\chi$ gradually approaches $\pi/2$ with the increase of $\theta$ and $N$.  This implies  the change of the probability distribution for the zero-energy state in Eq.~(\ref{zero}) from  the giant atom to the edge states of the SSH chains. In the inset of Fig.~\ref{Re_A1}(b), we display variation of $\chi$ with $\theta$ for $N=10$. The results show that $\chi$ is approached more quickly as $\pi/2$ when $N=10$ compared to $N=6$.

\begin{figure}[t]
  \centering
  \includegraphics[width=9cm]{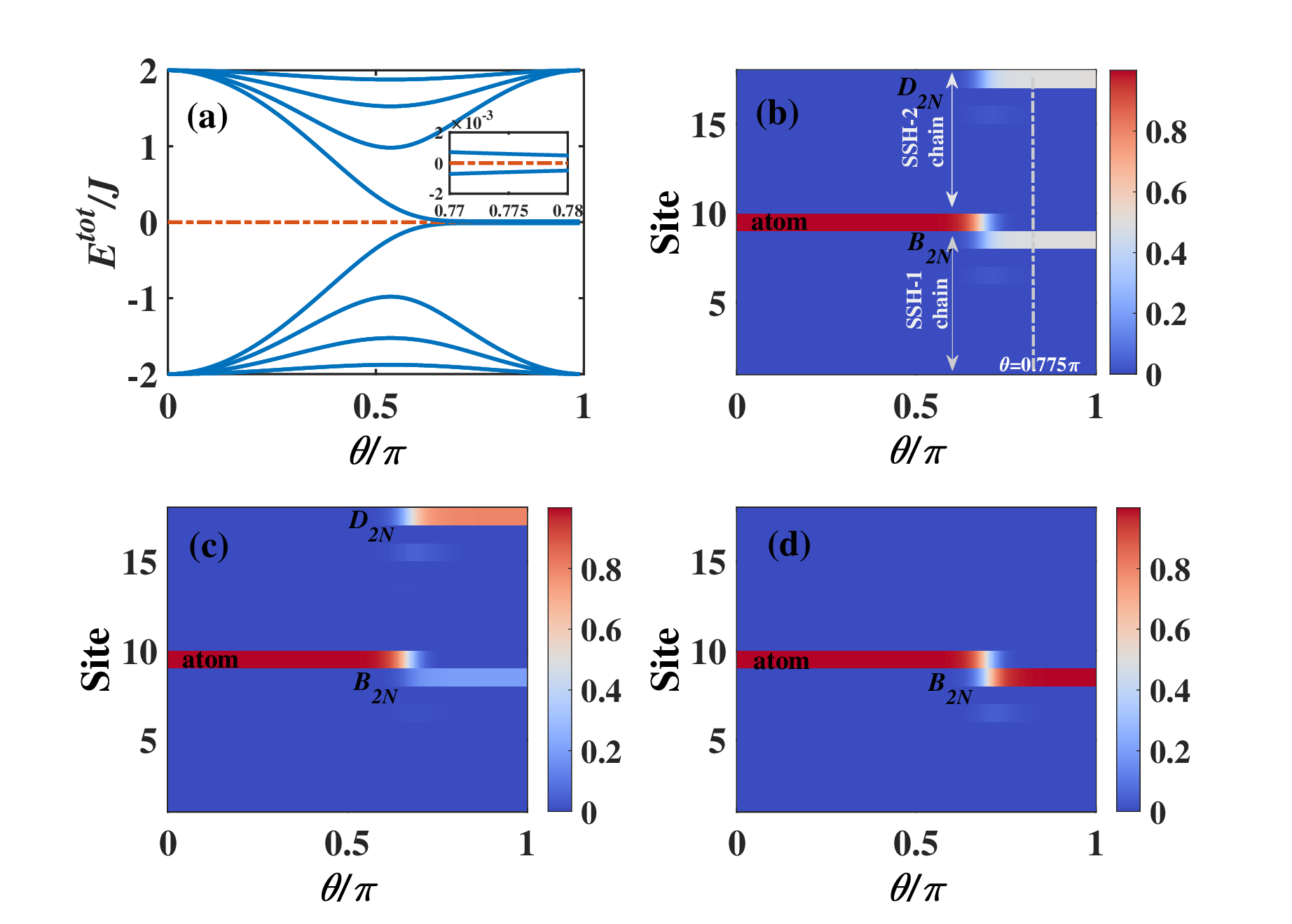}
  \caption{(a) The energy spectrum  of the total Hamiltonian in Eq.~(\ref{ee1}) with the variation of $\theta$ when the  giant atom is coupled to the sublattices $A_1$ and $C_1$. (b) The probability  distribution of the zero energy state of the panel (a) with orange dashed line on the sites  varies with $\theta$. [(c), (d)] The probability  distribution of the zero energy state on the sites varies with $\theta$ for the parameters  $g_1=0.01J$,  $g_2=0.02J$,  $p=1$, $q=1$~ (c) ; $g_1=g_2=0.01J$ and $p=1,q=4$~(d), respectively.  The  parameters are set to $N=4$ and $J=1$ for all figures.}\label{small_atom}
 \end{figure}

In order to prove the correctness of the five-state model, in Fig.~\ref{small_atom}(a), we plot the variation of energy spectrum $E^{tot}_{j}$ ($j=1,\cdots, 4N+1$) corresponding to the total Hamiltonian in Eq.~(\ref{ee1})  with the change of $\theta$ under the case that the giant atom is coupled to the sublattices $A_p$ and $C_q$.  It can be found that there exists a zero-energy state $|E^{tot}_{2N+1}\rangle$ within $\theta=[0,\pi]$.
In Fig.~\ref{small_atom}(b), we plot the probability distribution of $|E^{tot}_{2N+1}\rangle$ on the sites varying with $\theta$. We find that in the range $\theta \in [0, 0.5\pi]$, the probability distribution of $|E^{tot}_{2N+1}\rangle$ is concentrated on the giant atom, which is consistent with  Fig.~\ref{detuning_energy_evolution}(c), where there is no transition between giant atom and SSH chains.
With  changing  $\theta$ into topological phase,  the giant atom  interacts with SSH chains and  the component of   zero-energy state $|E^{tot}_{2N+1}\rangle$   gradually changes from the giant atom  to the rightmost sublattice $B_{2N}$ of SSH-1 chain and  sublattice $D_{2N}$ of SSH-2 chain with the same probability as $\theta$ increases. Furthermore, we can see that at $\theta=0.775\pi$ (white line)  the atomic excitations have been transferred to the edge states of two SSH chains, but the energy levels are still separated [see inset in Fig. ~\ref{small_atom}(a)], thus the requirement for adiabatic evolution is satisfied for Eq.~(\ref{La}).  We also plot the case for $g_2/g_1=2$ and $p=q=1$ in Fig.~\ref{small_atom}(c). The results also show the probability distribution of the two right edge states, depending on the modulus of the coupling strength between the atom and the SSH-1(2) chain, which  coincides with Figs.~\ref{energy_p}(c) and (d) obtained from  Eq.~(\ref{H_{eff}}). In Fig.~\ref{small_atom}(d), we consider that  the giant atom is coupled to different cells of two SSH chains such as $p=1, q=4$ with $g_2/g_1=1$. We can see that the probability distribution of the  zero-energy state transfers from the giant atom only to the right edge state of the SSH-$1$ chain and eventually to the rightmost sublattice $B_{2N}$.
With the same coupling strength $g_1=g_2$,  the transfer to the rightmost sublattices $B_{2N}$ and $D_{2N}$ of the SSH-1(2) chain can be realized by controlling the atomic coupling positions $p$ and $q$. Meanwhile, for $p=q$, we can determine the probability of information transfer to $B_{2N}$ and $D_{2N}$ by adjusting  $g_2/g_1$  the ratio of the coupling. Therefore, the zero-energy state obtained in subspace in  Eq.~(\ref{eq7}) is equivalent to the zero-energy state obtained in full  Hamiltonian in Eq.~(\ref{ee1}), which means that we can construct  adiabatic topological passages that allow the excitation of a giant atom to be delivered in a controlled way towards the ends of the two SSH chains.
\begin{figure}[t]
  \centering
  \includegraphics[width=9cm]{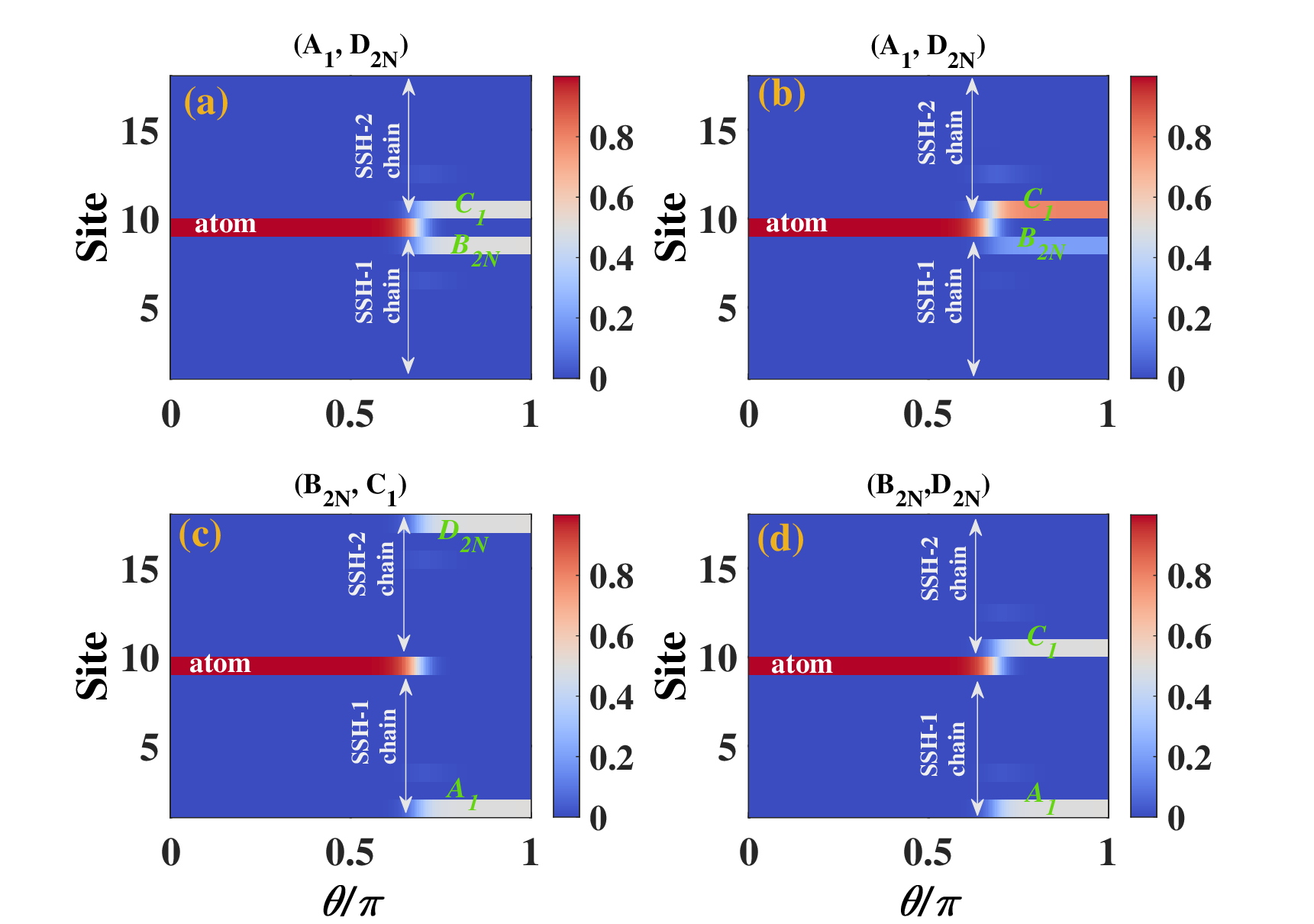}
  \caption{
  [(a)-(d)] The transfer of the atomic excitation for the coupling cases (a) and (b) $(a^\dag_1,d^\dag_{2N})$; (c) $(b^\dag_{2N},c^\dag_{1})$ and (d) $(b^\dag_{2N},d^\dag_{2N})$. The parameters are  $\Delta=0$ , $J=1$ and $g_1=g_2=0.01J$ except  $g_1=0.01J, g_2=0.02J$ for the panel (b). }\label{Re_A}
\end{figure}

 We also plot the component  distribution of zero energy state $|E^{tot}_{2N+1}\rangle$  corresponding to the other coupling cases $(a^\dag_1,d^\dag_{2N})$, $(b^\dag_{2N},c^\dag_{1})$  and $ (b^\dag_{2N},d^\dag_{2N})$.
As shown in Figs.~\ref{Re_A}(a) and (b), we consider the coupling of the giant atom to the two SSH chains via sublattices $A_1$ and $D_{2N}$ with the coupling strength $g_1=g_2=0.01J$ and  $g_1=0.01J, g_2=0.02J$, respectively.   It can be found that the zero-energy state probability distribution gradually transfers from the giant atom to the rightmost sublattice of the SSH-$1$ chain $B_{2N}$ and the leftmost sublattice of the SSH-$2$ chain $C_{1}$.  Meanwhile, the excitation occupancy in $B_{2N}$ and $C_{1}$ also depends on the square of the ratio $g_2/g_1$ of couplings.
For the remaining two coupling scenarios, we also display them in Figs.~\ref{Re_A}(c) and (d).
Therefore, we can conclude that when  giant  atom is coupled to the two SSH chains, there will exist a zero-energy state of the system, whose distribution is dominated by giant  atom when the  two SSH chains are in the trivial phase. With the adjustment of $\theta$, the two SSH chains enter the topological phase, and the distribution of the zero energy state will be gradually transferred to the end of the two SSH chains.
The end probabilities of the two SSH chains can be adjusted by adjusting the coupling strength of the atoms to the two SSH chains.
This means that we have established the adiabatic topological passages  assisted by  the zero energy state that can be used for controllable  quantum information transfer, where the  giant atom  acts as the  signal transmitter and the ends of the two SSH chains ($A_1$,$B_{2N}$,$C_1$,$D_{2N}$) act as information receivers.  The information transfer to the leftmost or rightmost end of the two SSH chains is achieved by choosing  the coupling points of the giant atom to the SSH chains, and the final probability occupancy is controlled by the coupling ratio $g_2/g_1$.

\begin{figure}[t]
  \centering
  \includegraphics[width=9cm]{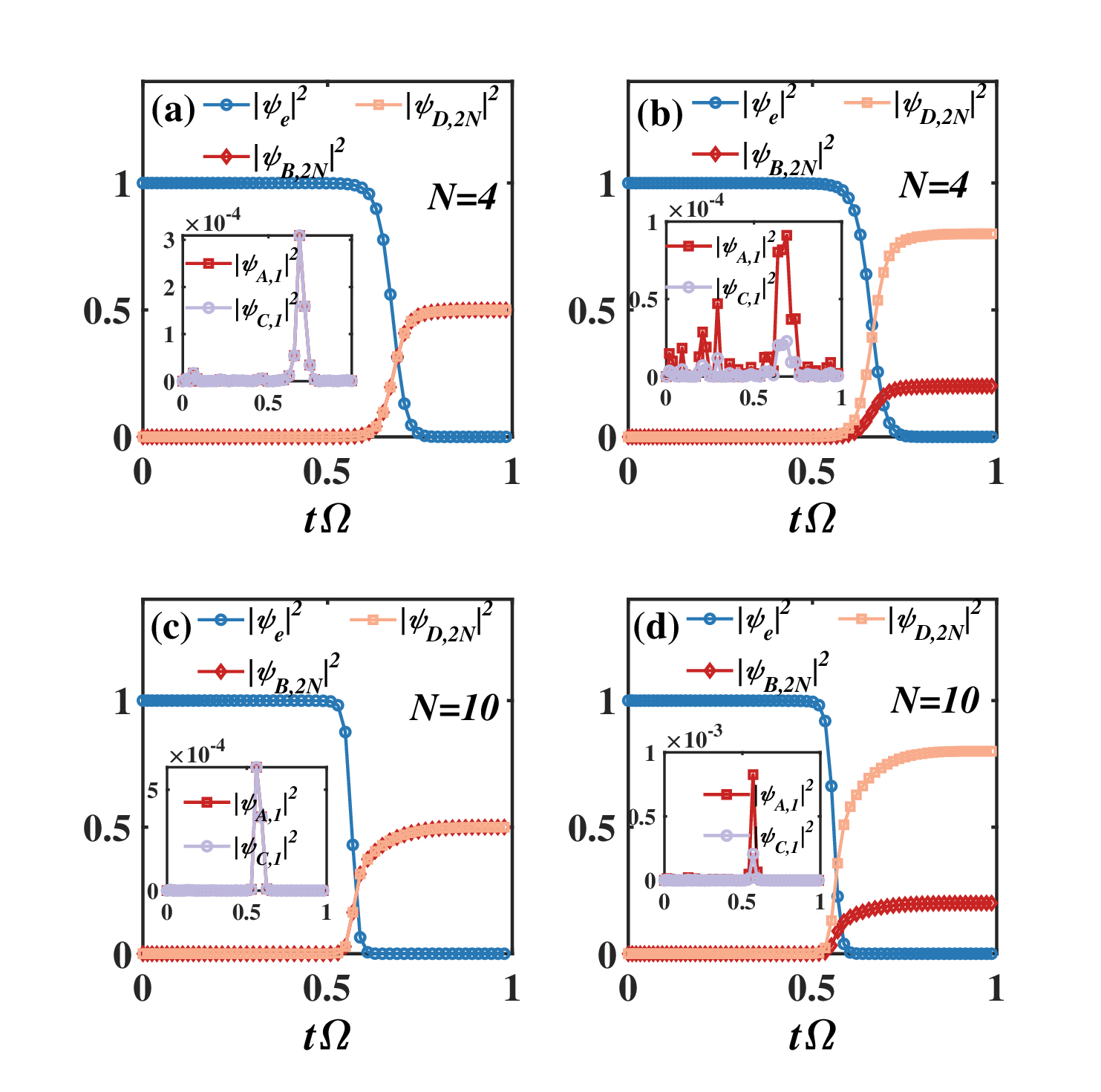}
  \caption{[(a)-(d)] The probability evolution of  the atomic excited state $|\psi_e|^2$, sublattice $A_1$ ($|\psi_{A,1}|^2$), sublattice $B_{2N}$ ($|\psi_{B,2N}|^2$),  sublattice $C_1$ ($|\psi_{C,1}|^2$) and  sublattice $D_{2N}$ ($|\psi_{D,2N}|^2$) versus the time for the giant atom coupled to the sublattice $A_1$ and $C_1$ with $\Omega=10^{-4}J$ and $g_1=g_2=0.01J$ for panels [(a), (c)], $g_1=0.01J, g_2=0.02J$ for panels [(b), (d)]. The  parameters are set to $J=1$, $N=4$ for panels [(a), (b)], $N=10$ for panels [(c), (d)]. }\label{evolution_ab}
 \end{figure}

We now confirm the validity of the adiabatic topological passage by directly employing the evolution results of the Hamiltonian~(\ref{ee1}) with slow time-varying rates, using $N=4$ as an example.  We consider the system is initially in the state $|\psi_i\rangle=\vert 0,\cdots,1_{\rm{atom}},\cdots,0\rangle$, in which the giant atom is in the excited state and the two topological chains are in the vacuum state.  We choose the two cases discussed in Fig.~\ref{small_atom} for $p=q=1$ with either $g_1=g_2=0.01J$ or $g_1=0.01J, g_2=0.02J$.  With Schr{\"o}dinger equation, we can derive the final state $|\psi_f\rangle$.
The results of the numerical simulations are displayed in Figs.~\ref{evolution_ab}(a) and (b), where $|\psi_{A,1}|^2$($|\psi_{C,1}|^2$) and $|\psi_{B,2N}|^2$ ($|\psi_{D,2N}|^2$) are the probabilities of  the leftmost and rightmost sublattices of the SSH-$1$ chain (SSH-2 chain), respectively.
We can see the giant atomic excitation gradually transfers to the rightmost sublattice of  the SSH-$1$ chain $B_{2N}$ and the leftmost sublattice of the SSH-$2$ chain $D_{2N}$ by adiabatic evolution with $\Omega=0.0001J$.  Furthermore,  the results obtained from time-dependent Hamiltonian~(\ref{ee1}) are consistent with Figs.~\ref{energy_p}(c) and (d) and Figs.~\ref{small_atom}(b) and (d), which proves that our previous discussions in the subspace are correct, namely  the adiabatic  topological passages exist.
We also study the adiabatic evolution for $N=10$,  the results are shown in Figs.~\ref{evolution_ab}(c) and (d). It can be found that for $N=10$, it is still possible to realize the transfer of atomic excitations to the ends of two SSH chains.

 \section{ Imperfections}\label{III}
\begin{figure}[t]
  \centering
  \includegraphics[width=9cm]{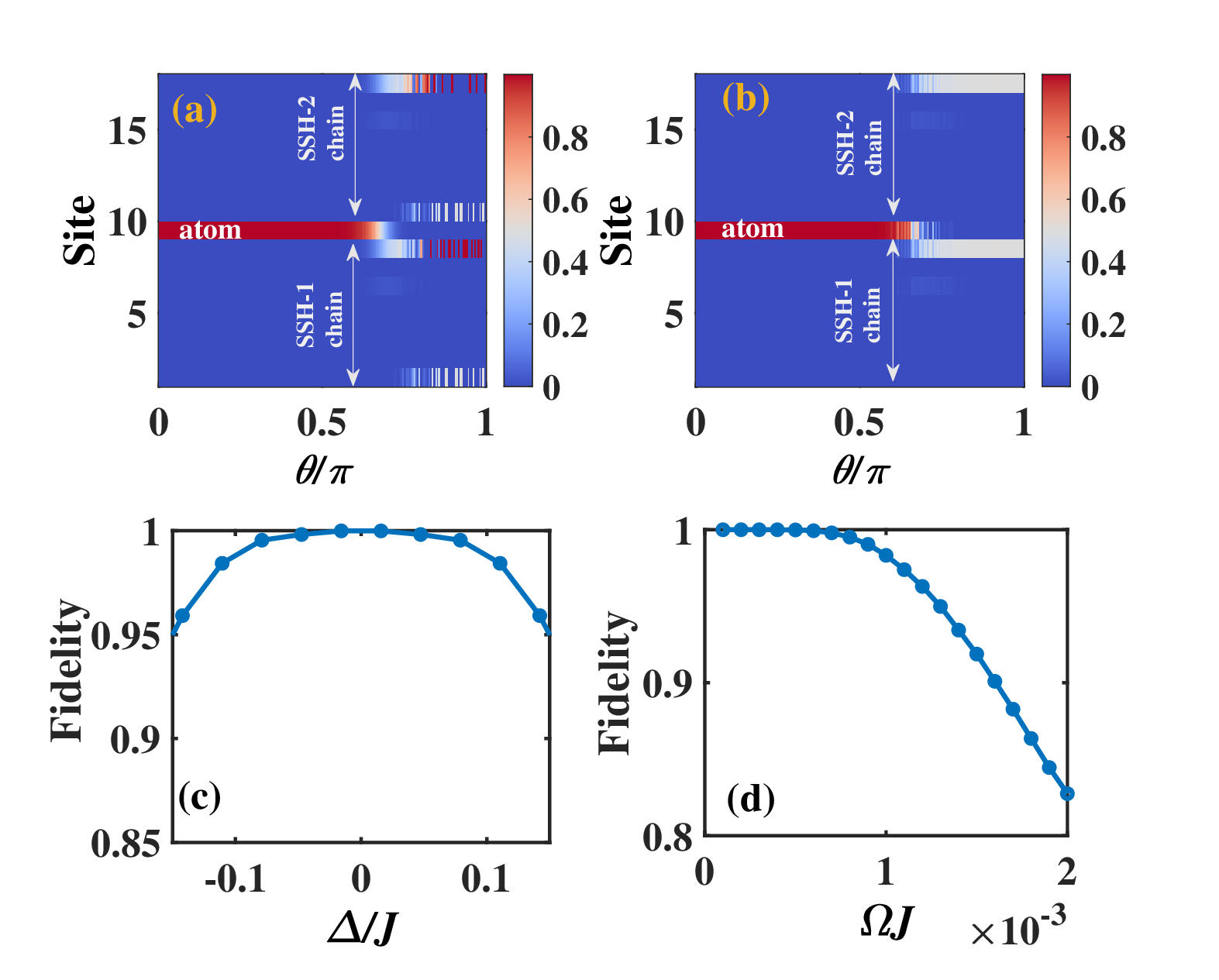}
  \caption{The probability distribution of zero-energy state on the sites varies with $\theta$ for the disorders of the lattice frequencies  within the region $\delta\omega_j\in[-0.001J, 0.001J]$ in (a) and the disorder couplings between lattices with $\eta_j\in[-0.001J, 0.001J]$ in (b), respectively.
(c) and (d) At time $t_f=\pi/\Omega$, the fidelity varies with $\Delta$ and $\Omega$, respectively.
 The  parameters are set to $J=1$, $N=4$ for panels [(a-d)], $\Delta=0$ for panels [(a),(b), (d)],  $\Omega=10^{-4}$ for panel (c). }\label{imperfertion}
 \end{figure}

 In this section, we discuss the effect of certain imperfections on the adiabatic topological passages.  We first discuss the implications of SSH chain-coupled disorder for the adiabatic topological passage.  Here there will be two types of disorders.  One comes from the disorders of the lattice frequencies, e.g.,  $(\omega_o+\delta\omega_j) a_j^\dag a_j$ and $(\omega_o+\delta\omega_j) b_j^\dag b_j$,  another one is from the disorder couplings between lattices, e.g.,  $(J_1+\eta_j) a_j^\dag b_j$ and $(J_2+\eta_j) a_{j+1}^\dag b_j$. We consider $\delta\omega_j$ and $\eta_j$ to be random numbers in the range $[-\xi,\xi]$,  where $\xi$ is the disorder strength. In Figs.~\ref{imperfertion}(a) and (b),  we plot the effect of  frequency disorder with $\xi=0.001J$  and coupling disorder with $\xi=0.1J$  on the probability distribution of zero-energy state, respectively. We can see that the adiabatic  topological passage is very sensitive to the frequency disorder and robust to the coupling disorder.  This is because the edge states of SSH chains are not affected by coupling disorder as long as the disorder strength does not cause a topological phase transition, but are extremely sensitive to frequency disorder.

 In Sec.~\ref{II}, we study the resonant interaction between the giant atom and SSH chains with $\Delta=0$. However, the non-resonant interaction with $\Delta\neq0$ is also very important. Let us now consider a target state $|\psi_t\rangle=(\vert 0,\cdots,1,0_{\rm{atom}},0,\cdots,0,0\rangle+\vert 0,\cdots,0,0_{\rm{atom}},0,\cdots,0,1\rangle)/\sqrt{2}$, which refers to the superposition of the rightmost lattices  $B_{2N}$ and  $D_{2N}$ at their excited states for the two SSH chains. This state can be obtained with an evolution time $t_{f}=\pi/\Omega$ for the resonant interaction between the giant atom and the SSH chains when the giant atom is initially in the excited state and the two topological chains are in their vacuum states. However, in the non-resonant case, the state $|\psi_f\rangle$ at the time  $t_{f}=\pi/\Omega$ is different from the state $|\psi_t\rangle$. The variation of the fidelity $F=\langle\psi_t|\psi_f\rangle$ with $\Delta$ is plotted in Fig.~\ref{imperfertion}(c).
 We can see that for $|\Delta|<0.15J$,  the fidelity is greater than $95\%$, but  the fidelity decreases rapidly as the increase of $\Delta$.  This is due to the fact that when the detuning $\Delta$  is large, the giant atom no longer resonates with the band gap states of the SSH chains, see Eqs.~(\ref{e6}) and  (\ref{e16}),   the fidelity decreases rapidly. Thus,  the adiabatic  topological passage is immune to weak coupling disorders and very small frequency mismatch between the giant atom and the edge states of the SSH chains.

\section{discussion}\label{IV}

We now  discuss the experimental feasibility. In recent years, based on the design flexibility and parameter controllability of the system structure~\cite{Gu2017,10.1063/1.5089550,RevModPhys.93.025005,doi:10.1126/science.1069372}, superconducting quantum circuits have attracted much attention and become into a well-established platform for the study of quantum simulation~ \cite{doi:10.1126/science.1177838,PhysRevX.7.011016,Noh2017} and quantum information processing~\cite{Daley2022,PhysRevA.75.032329,Mirhosseini2019}.
Recently, the works~\cite{tao2023,gu2017topologicaledgestatespumping,PhysRevX.11.011015,Bello2019,PhysRevLett.126.063601} have theoretically and experimentally shown that superconducting quantum circuits can be used to simulate the SSH model, where each cell of the SSH chain can be simulated by either two~\text{LC} resonators or qubits. Moreover, the periodic modulation of the coupling can also be realized~\cite{Houck2012,Schmidt2013}.  Therefore, the SSH chains may be realized by superconducting quantum circuits.

The superconducting qubit circuits acting as the giant atoms have been experimentally demonstrated~\cite{Kannan2020,Andersson2019,PhysRevX.13.021039}.  We know that the giant atom
 inevitably interacts with the environment which results in the loss of the coherence, thus the adiabatic evolution time $t_f$ required to complete the transfer of atomic excitations to the ends of the two SSH chains must be less than the atomic decoherence time.  Recently, coherence times from several tens of microseconds to milliseconds ($t\approx10^{-3}$s) for superconducting qubits have been demonstrated based on the state-of-the-art experimental systems~\cite{PhysRevLett.130.267001}. In~\cite{PhysRevX.11.011015}, the coupling strength of the SSH chain can be about $J\approx 10^8$Hz, and thus the time to complete the adiabatic state transfer in our system is about $t_f=\pi/(\Omega J)=10^{-4}$s, which is less than the decoherence time of the superconducting qubits. We also plot the fidelity defined in Sec.~\ref{III} at the time $t_f=\pi/\Omega$ varying with $\Omega$  as shown in Fig.~\ref{imperfertion}(d). It can be seen that  high fidelity can still be achieved even with a low evolution speed $\Omega J=10^{-3}$ corresponding to the evolution time $t_f=10^{-5}$s. Thus  our proposal may be observed in the experiments.

\section{conclusion}\label{V}
We propose an approach to construct an adiabatic topological passage by coupling a giant atom to two SSH chains. The adiabatic topological passage allows the excitation of the giant atom to be delivered in a controllable way  to both ends of the two SSH chains. When two finite SSH chains are in the topological phase and  the giant atom resonates with the edge states of the SSH chain, the system is reduced to  the coupling of the giant atom to the edge states of the two SSH chains. Then the system is equivalent  to a five-state model, which includes the edge states of the two SSH chains as well as the giant atom. In this case, we find the presence of dark states in the subspace, which can be worked as a  adiabatic topological passage.
In this way, we can  use adiabatic evolution to transfer the excitation of the giant atom to one end of the two SSH chains in a fully controllable way.
In addition, we also show the robustness of the adiabatic topological passage  to the coupling disorders in SSH chains and the frequency mismatch between the giant atom and edge states of the SSH chains. Our study may be realized by the superconducting quantum circuits.

\section{acknowledgments}
This work was supported by the National Natural Science Foundation of China Grants No. 12374483, No.12274053  and No. 92365209.
\begin{appendix}

\end{appendix}
\bibliography{seven}

\begin{thebibliography}{80}%
\makeatletter
\providecommand \@ifxundefined [1]{%
 \@ifx{#1\undefined}
}%
\providecommand \@ifnum [1]{%
 \ifnum #1\expandafter \@firstoftwo
 \else \expandafter \@secondoftwo
 \fi
}%
\providecommand \@ifx [1]{%
 \ifx #1\expandafter \@firstoftwo
 \else \expandafter \@secondoftwo
 \fi
}%
\providecommand \natexlab [1]{#1}%
\providecommand \enquote  [1]{``#1''}%
\providecommand \bibnamefont  [1]{#1}%
\providecommand \bibfnamefont [1]{#1}%
\providecommand \citenamefont [1]{#1}%
\providecommand \href@noop [0]{\@secondoftwo}%
\providecommand \href [0]{\begingroup \@sanitize@url \@href}%
\providecommand \@href[1]{\@@startlink{#1}\@@href}%
\providecommand \@@href[1]{\endgroup#1\@@endlink}%
\providecommand \@sanitize@url [0]{\catcode `\\12\catcode `\$12\catcode
  `\&12\catcode `\#12\catcode `\^12\catcode `\_12\catcode `\%12\relax}%
\providecommand \@@startlink[1]{}%
\providecommand \@@endlink[0]{}%
\providecommand \url  [0]{\begingroup\@sanitize@url \@url }%
\providecommand \@url [1]{\endgroup\@href {#1}{\urlprefix }}%
\providecommand \urlprefix  [0]{URL }%
\providecommand \Eprint [0]{\href }%
\providecommand \doibase [0]{http://dx.doi.org/}%
\providecommand \selectlanguage [0]{\@gobble}%
\providecommand \bibinfo  [0]{\@secondoftwo}%
\providecommand \bibfield  [0]{\@secondoftwo}%
\providecommand \translation [1]{[#1]}%
\providecommand \BibitemOpen [0]{}%
\providecommand \bibitemStop [0]{}%
\providecommand \bibitemNoStop [0]{.\EOS\space}%
\providecommand \EOS [0]{\spacefactor3000\relax}%
\providecommand \BibitemShut  [1]{\csname bibitem#1\endcsname}%
\let\auto@bib@innerbib\@empty
\bibitem [{\citenamefont {Kimble}(2008)}]{Kimble2008}%
  \BibitemOpen
  \bibfield  {author} {\bibinfo {author} {\bibfnamefont {H.~J.}\ \bibnamefont
  {Kimble}},\ }\href {\doibase 10.1038/nature07127} {\bibfield  {journal}
  {\bibinfo  {journal} {Nature}\ }\textbf {\bibinfo {volume} {453}},\ \bibinfo
  {pages} {1023} (\bibinfo {year} {2008})}\BibitemShut {NoStop}%
\bibitem [{\citenamefont {Cirac}\ \emph {et~al.}(1997)\citenamefont {Cirac},
  \citenamefont {Zoller}, \citenamefont {Kimble},\ and\ \citenamefont
  {Mabuchi}}]{PhysRevLett.78.3221}%
  \BibitemOpen
  \bibfield  {author} {\bibinfo {author} {\bibfnamefont {J.~I.}\ \bibnamefont
  {Cirac}}, \bibinfo {author} {\bibfnamefont {P.}~\bibnamefont {Zoller}},
  \bibinfo {author} {\bibfnamefont {H.~J.}\ \bibnamefont {Kimble}}, \ and\
  \bibinfo {author} {\bibfnamefont {H.}~\bibnamefont {Mabuchi}},\ }\href
  {\doibase 10.1103/PhysRevLett.78.3221} {\bibfield  {journal} {\bibinfo
  {journal} {Phys. Rev. Lett.}\ }\textbf {\bibinfo {volume} {78}},\ \bibinfo
  {pages} {3221} (\bibinfo {year} {1997})}\BibitemShut {NoStop}%
\bibitem [{\citenamefont {Chen}\ \emph {et~al.}(2011)\citenamefont {Chen},
  \citenamefont {Nascimb\`ene}, \citenamefont {Aidelsburger}, \citenamefont
  {Atala}, \citenamefont {Trotzky},\ and\ \citenamefont
  {Bloch}}]{PhysRevLett.107.210405}%
  \BibitemOpen
  \bibfield  {author} {\bibinfo {author} {\bibfnamefont {Y.-A.}\ \bibnamefont
  {Chen}}, \bibinfo {author} {\bibfnamefont {S.}~\bibnamefont {Nascimb\`ene}},
  \bibinfo {author} {\bibfnamefont {M.}~\bibnamefont {Aidelsburger}}, \bibinfo
  {author} {\bibfnamefont {M.}~\bibnamefont {Atala}}, \bibinfo {author}
  {\bibfnamefont {S.}~\bibnamefont {Trotzky}}, \ and\ \bibinfo {author}
  {\bibfnamefont {I.}~\bibnamefont {Bloch}},\ }\href {\doibase
  10.1103/PhysRevLett.107.210405} {\bibfield  {journal} {\bibinfo  {journal}
  {Phys. Rev. Lett.}\ }\textbf {\bibinfo {volume} {107}},\ \bibinfo {pages}
  {210405} (\bibinfo {year} {2011})}\BibitemShut {NoStop}%
\bibitem [{\citenamefont {Reiserer}\ and\ \citenamefont
  {Rempe}(2015)}]{RevModPhys.87.1379}%
  \BibitemOpen
  \bibfield  {author} {\bibinfo {author} {\bibfnamefont {A.}~\bibnamefont
  {Reiserer}}\ and\ \bibinfo {author} {\bibfnamefont {G.}~\bibnamefont
  {Rempe}},\ }\href {\doibase 10.1103/RevModPhys.87.1379} {\bibfield  {journal}
  {\bibinfo  {journal} {Rev. Mod. Phys.}\ }\textbf {\bibinfo {volume} {87}},\
  \bibinfo {pages} {1379} (\bibinfo {year} {2015})}\BibitemShut {NoStop}%
\bibitem [{\citenamefont {Ritter}\ \emph {et~al.}(2012)\citenamefont {Ritter},
  \citenamefont {N{\"o}lleke}, \citenamefont {Hahn}, \citenamefont {Reiserer},
  \citenamefont {Neuzner}, \citenamefont {Uphoff}, \citenamefont {M{\"u}cke},
  \citenamefont {Figueroa}, \citenamefont {Bochmann},\ and\ \citenamefont
  {Rempe}}]{Ritter2012}%
  \BibitemOpen
  \bibfield  {author} {\bibinfo {author} {\bibfnamefont {S.}~\bibnamefont
  {Ritter}}, \bibinfo {author} {\bibfnamefont {C.}~\bibnamefont {N{\"o}lleke}},
  \bibinfo {author} {\bibfnamefont {C.}~\bibnamefont {Hahn}}, \bibinfo {author}
  {\bibfnamefont {A.}~\bibnamefont {Reiserer}}, \bibinfo {author}
  {\bibfnamefont {A.}~\bibnamefont {Neuzner}}, \bibinfo {author} {\bibfnamefont
  {M.}~\bibnamefont {Uphoff}}, \bibinfo {author} {\bibfnamefont
  {M.}~\bibnamefont {M{\"u}cke}}, \bibinfo {author} {\bibfnamefont
  {E.}~\bibnamefont {Figueroa}}, \bibinfo {author} {\bibfnamefont
  {J.}~\bibnamefont {Bochmann}}, \ and\ \bibinfo {author} {\bibfnamefont
  {G.}~\bibnamefont {Rempe}},\ }\href {\doibase 10.1038/nature11023} {\bibfield
   {journal} {\bibinfo  {journal} {Nature}\ }\textbf {\bibinfo {volume}
  {484}},\ \bibinfo {pages} {195} (\bibinfo {year} {2012})}\BibitemShut
  {NoStop}%
\bibitem [{\citenamefont {Vermersch}\ \emph {et~al.}(2017)\citenamefont
  {Vermersch}, \citenamefont {Guimond}, \citenamefont {Pichler},\ and\
  \citenamefont {Zoller}}]{PhysRevLett.118.133601}%
  \BibitemOpen
  \bibfield  {author} {\bibinfo {author} {\bibfnamefont {B.}~\bibnamefont
  {Vermersch}}, \bibinfo {author} {\bibfnamefont {P.-O.}\ \bibnamefont
  {Guimond}}, \bibinfo {author} {\bibfnamefont {H.}~\bibnamefont {Pichler}}, \
  and\ \bibinfo {author} {\bibfnamefont {P.}~\bibnamefont {Zoller}},\ }\href
  {\doibase 10.1103/PhysRevLett.118.133601} {\bibfield  {journal} {\bibinfo
  {journal} {Phys. Rev. Lett.}\ }\textbf {\bibinfo {volume} {118}},\ \bibinfo
  {pages} {133601} (\bibinfo {year} {2017})}\BibitemShut {NoStop}%
\bibitem [{\citenamefont {Kuzyk}\ and\ \citenamefont
  {Wang}(2018)}]{PhysRevX.8.041027}%
  \BibitemOpen
  \bibfield  {author} {\bibinfo {author} {\bibfnamefont {M.~C.}\ \bibnamefont
  {Kuzyk}}\ and\ \bibinfo {author} {\bibfnamefont {H.}~\bibnamefont {Wang}},\
  }\href {\doibase 10.1103/PhysRevX.8.041027} {\bibfield  {journal} {\bibinfo
  {journal} {Phys. Rev. X}\ }\textbf {\bibinfo {volume} {8}},\ \bibinfo {pages}
  {041027} (\bibinfo {year} {2018})}\BibitemShut {NoStop}%
\bibitem [{\citenamefont {Axline}\ \emph {et~al.}(2018)\citenamefont {Axline},
  \citenamefont {Burkhart}, \citenamefont {Pfaff}, \citenamefont {Zhang},
  \citenamefont {Chou}, \citenamefont {Campagne-Ibarcq}, \citenamefont
  {Reinhold}, \citenamefont {Frunzio}, \citenamefont {Girvin}, \citenamefont
  {Jiang}, \citenamefont {Devoret},\ and\ \citenamefont
  {Schoelkopf}}]{Axline2018}%
  \BibitemOpen
  \bibfield  {author} {\bibinfo {author} {\bibfnamefont {C.~J.}\ \bibnamefont
  {Axline}}, \bibinfo {author} {\bibfnamefont {L.~D.}\ \bibnamefont
  {Burkhart}}, \bibinfo {author} {\bibfnamefont {W.}~\bibnamefont {Pfaff}},
  \bibinfo {author} {\bibfnamefont {M.}~\bibnamefont {Zhang}}, \bibinfo
  {author} {\bibfnamefont {K.}~\bibnamefont {Chou}}, \bibinfo {author}
  {\bibfnamefont {P.}~\bibnamefont {Campagne-Ibarcq}}, \bibinfo {author}
  {\bibfnamefont {P.}~\bibnamefont {Reinhold}}, \bibinfo {author}
  {\bibfnamefont {L.}~\bibnamefont {Frunzio}}, \bibinfo {author} {\bibfnamefont
  {S.~M.}\ \bibnamefont {Girvin}}, \bibinfo {author} {\bibfnamefont
  {L.}~\bibnamefont {Jiang}}, \bibinfo {author} {\bibfnamefont {M.~H.}\
  \bibnamefont {Devoret}}, \ and\ \bibinfo {author} {\bibfnamefont {R.~J.}\
  \bibnamefont {Schoelkopf}},\ }\href {\doibase 10.1038/s41567-018-0115-y}
  {\bibfield  {journal} {\bibinfo  {journal} {Nature Physics}\ }\textbf
  {\bibinfo {volume} {14}},\ \bibinfo {pages} {705} (\bibinfo {year}
  {2018})}\BibitemShut {NoStop}%
\bibitem [{\citenamefont {Mei}\ \emph {et~al.}(2018)\citenamefont {Mei},
  \citenamefont {Chen}, \citenamefont {Tian}, \citenamefont {Zhu},\ and\
  \citenamefont {Jia}}]{PhysRevA.98.012331}%
  \BibitemOpen
  \bibfield  {author} {\bibinfo {author} {\bibfnamefont {F.}~\bibnamefont
  {Mei}}, \bibinfo {author} {\bibfnamefont {G.}~\bibnamefont {Chen}}, \bibinfo
  {author} {\bibfnamefont {L.}~\bibnamefont {Tian}}, \bibinfo {author}
  {\bibfnamefont {S.-L.}\ \bibnamefont {Zhu}}, \ and\ \bibinfo {author}
  {\bibfnamefont {S.}~\bibnamefont {Jia}},\ }\href {\doibase
  10.1103/PhysRevA.98.012331} {\bibfield  {journal} {\bibinfo  {journal} {Phys.
  Rev. A}\ }\textbf {\bibinfo {volume} {98}},\ \bibinfo {pages} {012331}
  (\bibinfo {year} {2018})}\BibitemShut {NoStop}%
\bibitem [{\citenamefont {Petrosyan}\ and\ \citenamefont
  {Lambropoulos}(2006)}]{PETROSYAN2006419}%
  \BibitemOpen
  \bibfield  {author} {\bibinfo {author} {\bibfnamefont {D.}~\bibnamefont
  {Petrosyan}}\ and\ \bibinfo {author} {\bibfnamefont {P.}~\bibnamefont
  {Lambropoulos}},\ }\href {\doibase
  https://doi.org/10.1016/j.optcom.2005.12.082} {\bibfield  {journal} {\bibinfo
   {journal} {Opt. Commun.}\ }\textbf {\bibinfo {volume} {264}},\ \bibinfo
  {pages} {419} (\bibinfo {year} {2006})},\ \bibinfo {note} {quantum Control of
  Light and Matter}\BibitemShut {NoStop}%
\bibitem [{\citenamefont {Vitanov}\ \emph {et~al.}(2017)\citenamefont
  {Vitanov}, \citenamefont {Rangelov}, \citenamefont {Shore},\ and\
  \citenamefont {Bergmann}}]{RevModPhys.89.015006}%
  \BibitemOpen
  \bibfield  {author} {\bibinfo {author} {\bibfnamefont {N.~V.}\ \bibnamefont
  {Vitanov}}, \bibinfo {author} {\bibfnamefont {A.~A.}\ \bibnamefont
  {Rangelov}}, \bibinfo {author} {\bibfnamefont {B.~W.}\ \bibnamefont {Shore}},
  \ and\ \bibinfo {author} {\bibfnamefont {K.}~\bibnamefont {Bergmann}},\
  }\href {\doibase 10.1103/RevModPhys.89.015006} {\bibfield  {journal}
  {\bibinfo  {journal} {Rev. Mod. Phys.}\ }\textbf {\bibinfo {volume} {89}},\
  \bibinfo {pages} {015006} (\bibinfo {year} {2017})}\BibitemShut {NoStop}%
\bibitem [{\citenamefont {Kuklinski}\ \emph {et~al.}(1989)\citenamefont
  {Kuklinski}, \citenamefont {Gaubatz}, \citenamefont {Hioe},\ and\
  \citenamefont {Bergmann}}]{PhysRevA.40.6741}%
  \BibitemOpen
  \bibfield  {author} {\bibinfo {author} {\bibfnamefont {J.~R.}\ \bibnamefont
  {Kuklinski}}, \bibinfo {author} {\bibfnamefont {U.}~\bibnamefont {Gaubatz}},
  \bibinfo {author} {\bibfnamefont {F.~T.}\ \bibnamefont {Hioe}}, \ and\
  \bibinfo {author} {\bibfnamefont {K.}~\bibnamefont {Bergmann}},\ }\href
  {\doibase 10.1103/PhysRevA.40.6741} {\bibfield  {journal} {\bibinfo
  {journal} {Phys. Rev. A}\ }\textbf {\bibinfo {volume} {40}},\ \bibinfo
  {pages} {6741} (\bibinfo {year} {1989})}\BibitemShut {NoStop}%
\bibitem [{\citenamefont {Bergmann}\ \emph {et~al.}(1998)\citenamefont
  {Bergmann}, \citenamefont {Theuer},\ and\ \citenamefont
  {Shore}}]{RevModPhys.70.1003}%
  \BibitemOpen
  \bibfield  {author} {\bibinfo {author} {\bibfnamefont {K.}~\bibnamefont
  {Bergmann}}, \bibinfo {author} {\bibfnamefont {H.}~\bibnamefont {Theuer}}, \
  and\ \bibinfo {author} {\bibfnamefont {B.~W.}\ \bibnamefont {Shore}},\ }\href
  {\doibase 10.1103/RevModPhys.70.1003} {\bibfield  {journal} {\bibinfo
  {journal} {Rev. Mod. Phys.}\ }\textbf {\bibinfo {volume} {70}},\ \bibinfo
  {pages} {1003} (\bibinfo {year} {1998})}\BibitemShut {NoStop}%
\bibitem [{\citenamefont {Chen}\ \emph {et~al.}(2013)\citenamefont {Chen},
  \citenamefont {Fan}, \citenamefont {Xu}, \citenamefont {Peng},\ and\
  \citenamefont {Zhang}}]{PhysRevA.88.022323}%
  \BibitemOpen
  \bibfield  {author} {\bibinfo {author} {\bibfnamefont {B.}~\bibnamefont
  {Chen}}, \bibinfo {author} {\bibfnamefont {W.}~\bibnamefont {Fan}}, \bibinfo
  {author} {\bibfnamefont {Y.}~\bibnamefont {Xu}}, \bibinfo {author}
  {\bibfnamefont {Y.-D.}\ \bibnamefont {Peng}}, \ and\ \bibinfo {author}
  {\bibfnamefont {H.-Y.}\ \bibnamefont {Zhang}},\ }\href {\doibase
  10.1103/PhysRevA.88.022323} {\bibfield  {journal} {\bibinfo  {journal} {Phys.
  Rev. A}\ }\textbf {\bibinfo {volume} {88}},\ \bibinfo {pages} {022323}
  (\bibinfo {year} {2013})}\BibitemShut {NoStop}%
\bibitem [{\citenamefont {Hasan}\ and\ \citenamefont
  {Kane}(2010)}]{RevModPhys.82.3045}%
  \BibitemOpen
  \bibfield  {author} {\bibinfo {author} {\bibfnamefont {M.~Z.}\ \bibnamefont
  {Hasan}}\ and\ \bibinfo {author} {\bibfnamefont {C.~L.}\ \bibnamefont
  {Kane}},\ }\href {\doibase 10.1103/RevModPhys.82.3045} {\bibfield  {journal}
  {\bibinfo  {journal} {Rev. Mod. Phys.}\ }\textbf {\bibinfo {volume} {82}},\
  \bibinfo {pages} {3045} (\bibinfo {year} {2010})}\BibitemShut {NoStop}%
\bibitem [{\citenamefont {Wray}\ \emph {et~al.}(2010)\citenamefont {Wray},
  \citenamefont {Xu}, \citenamefont {Xia}, \citenamefont {Hor}, \citenamefont
  {Qian}, \citenamefont {Fedorov}, \citenamefont {Lin}, \citenamefont {Bansil},
  \citenamefont {Cava},\ and\ \citenamefont {Hasan}}]{Wray2010}%
  \BibitemOpen
  \bibfield  {author} {\bibinfo {author} {\bibfnamefont {L.~A.}\ \bibnamefont
  {Wray}}, \bibinfo {author} {\bibfnamefont {S.~Y.}\ \bibnamefont {Xu}},
  \bibinfo {author} {\bibfnamefont {Y.}~\bibnamefont {Xia}}, \bibinfo {author}
  {\bibfnamefont {Y.~S.}\ \bibnamefont {Hor}}, \bibinfo {author} {\bibfnamefont
  {D.}~\bibnamefont {Qian}}, \bibinfo {author} {\bibfnamefont {A.~V.}\
  \bibnamefont {Fedorov}}, \bibinfo {author} {\bibfnamefont {H.}~\bibnamefont
  {Lin}}, \bibinfo {author} {\bibfnamefont {A.}~\bibnamefont {Bansil}},
  \bibinfo {author} {\bibfnamefont {R.~J.}\ \bibnamefont {Cava}}, \ and\
  \bibinfo {author} {\bibfnamefont {M.~Z.}\ \bibnamefont {Hasan}},\ }\href
  {\doibase 10.1038/nphys1762} {\bibfield  {journal} {\bibinfo  {journal} {Nat.
  Phys.}\ }\textbf {\bibinfo {volume} {6}},\ \bibinfo {pages} {855} (\bibinfo
  {year} {2010})}\BibitemShut {NoStop}%
\bibitem [{\citenamefont {Shalaev}\ \emph {et~al.}(2019)\citenamefont
  {Shalaev}, \citenamefont {Walasik}, \citenamefont {Tsukernik}, \citenamefont
  {Xu},\ and\ \citenamefont {Litchinitser}}]{Shalaev2019}%
  \BibitemOpen
  \bibfield  {author} {\bibinfo {author} {\bibfnamefont {M.~I.}\ \bibnamefont
  {Shalaev}}, \bibinfo {author} {\bibfnamefont {W.}~\bibnamefont {Walasik}},
  \bibinfo {author} {\bibfnamefont {A.}~\bibnamefont {Tsukernik}}, \bibinfo
  {author} {\bibfnamefont {Y.}~\bibnamefont {Xu}}, \ and\ \bibinfo {author}
  {\bibfnamefont {N.~M.}\ \bibnamefont {Litchinitser}},\ }\href {\doibase
  10.1038/s41565-018-0297-6} {\bibfield  {journal} {\bibinfo  {journal} {Nat.
  Nanotechnol.}\ }\textbf {\bibinfo {volume} {14}},\ \bibinfo {pages} {31}
  (\bibinfo {year} {2019})}\BibitemShut {NoStop}%
\bibitem [{\citenamefont {Ozawa}\ \emph {et~al.}(2019)\citenamefont {Ozawa},
  \citenamefont {Price}, \citenamefont {Amo}, \citenamefont {Goldman},
  \citenamefont {Hafezi}, \citenamefont {Lu}, \citenamefont {Rechtsman},
  \citenamefont {Schuster}, \citenamefont {Simon}, \citenamefont {Zilberberg},\
  and\ \citenamefont {Carusotto}}]{RevModPhys.91.015006}%
  \BibitemOpen
  \bibfield  {author} {\bibinfo {author} {\bibfnamefont {T.}~\bibnamefont
  {Ozawa}}, \bibinfo {author} {\bibfnamefont {H.~M.}\ \bibnamefont {Price}},
  \bibinfo {author} {\bibfnamefont {A.}~\bibnamefont {Amo}}, \bibinfo {author}
  {\bibfnamefont {N.}~\bibnamefont {Goldman}}, \bibinfo {author} {\bibfnamefont
  {M.}~\bibnamefont {Hafezi}}, \bibinfo {author} {\bibfnamefont
  {L.}~\bibnamefont {Lu}}, \bibinfo {author} {\bibfnamefont {M.~C.}\
  \bibnamefont {Rechtsman}}, \bibinfo {author} {\bibfnamefont {D.}~\bibnamefont
  {Schuster}}, \bibinfo {author} {\bibfnamefont {J.}~\bibnamefont {Simon}},
  \bibinfo {author} {\bibfnamefont {O.}~\bibnamefont {Zilberberg}}, \ and\
  \bibinfo {author} {\bibfnamefont {I.}~\bibnamefont {Carusotto}},\ }\href
  {\doibase 10.1103/RevModPhys.91.015006} {\bibfield  {journal} {\bibinfo
  {journal} {Rev. Mod. Phys.}\ }\textbf {\bibinfo {volume} {91}},\ \bibinfo
  {pages} {015006} (\bibinfo {year} {2019})}\BibitemShut {NoStop}%
\bibitem [{\citenamefont {Thouless}(1983)}]{PhysRevB.27.6083}%
  \BibitemOpen
  \bibfield  {author} {\bibinfo {author} {\bibfnamefont {D.~J.}\ \bibnamefont
  {Thouless}},\ }\href {\doibase 10.1103/PhysRevB.27.6083} {\bibfield
  {journal} {\bibinfo  {journal} {Phys. Rev. B}\ }\textbf {\bibinfo {volume}
  {27}},\ \bibinfo {pages} {6083} (\bibinfo {year} {1983})}\BibitemShut
  {NoStop}%
\bibitem [{\citenamefont {Kraus}\ \emph {et~al.}(2012)\citenamefont {Kraus},
  \citenamefont {Lahini}, \citenamefont {Ringel}, \citenamefont {Verbin},\ and\
  \citenamefont {Zilberberg}}]{PhysRevLett.109.106402}%
  \BibitemOpen
  \bibfield  {author} {\bibinfo {author} {\bibfnamefont {Y.~E.}\ \bibnamefont
  {Kraus}}, \bibinfo {author} {\bibfnamefont {Y.}~\bibnamefont {Lahini}},
  \bibinfo {author} {\bibfnamefont {Z.}~\bibnamefont {Ringel}}, \bibinfo
  {author} {\bibfnamefont {M.}~\bibnamefont {Verbin}}, \ and\ \bibinfo {author}
  {\bibfnamefont {O.}~\bibnamefont {Zilberberg}},\ }\href {\doibase
  10.1103/PhysRevLett.109.106402} {\bibfield  {journal} {\bibinfo  {journal}
  {Phys. Rev. Lett.}\ }\textbf {\bibinfo {volume} {109}},\ \bibinfo {pages}
  {106402} (\bibinfo {year} {2012})}\BibitemShut {NoStop}%
\bibitem [{\citenamefont {Longhi}\ \emph {et~al.}(2019)\citenamefont {Longhi},
  \citenamefont {Giorgi},\ and\ \citenamefont {Zambrini}}]{Longhi2019}%
  \BibitemOpen
  \bibfield  {author} {\bibinfo {author} {\bibfnamefont {S.}~\bibnamefont
  {Longhi}}, \bibinfo {author} {\bibfnamefont {G.~L.}\ \bibnamefont {Giorgi}},
  \ and\ \bibinfo {author} {\bibfnamefont {R.}~\bibnamefont {Zambrini}},\
  }\href {\doibase https://doi.org/10.1002/qute.201800090} {\bibfield
  {journal} {\bibinfo  {journal} {Adv. Quantum Technol.}\ }\textbf {\bibinfo
  {volume} {2}},\ \bibinfo {pages} {1800090} (\bibinfo {year}
  {2019})}\BibitemShut {NoStop}%
\bibitem [{\citenamefont {Palaiodimopoulos}\ \emph {et~al.}(2021)\citenamefont
  {Palaiodimopoulos}, \citenamefont {Brouzos}, \citenamefont {Diakonos},\ and\
  \citenamefont {Theocharis}}]{PhysRevA.103.052409}%
  \BibitemOpen
  \bibfield  {author} {\bibinfo {author} {\bibfnamefont {N.~E.}\ \bibnamefont
  {Palaiodimopoulos}}, \bibinfo {author} {\bibfnamefont {I.}~\bibnamefont
  {Brouzos}}, \bibinfo {author} {\bibfnamefont {F.~K.}\ \bibnamefont
  {Diakonos}}, \ and\ \bibinfo {author} {\bibfnamefont {G.}~\bibnamefont
  {Theocharis}},\ }\href {\doibase 10.1103/PhysRevA.103.052409} {\bibfield
  {journal} {\bibinfo  {journal} {Phys. Rev. A}\ }\textbf {\bibinfo {volume}
  {103}},\ \bibinfo {pages} {052409} (\bibinfo {year} {2021})}\BibitemShut
  {NoStop}%
\bibitem [{\citenamefont {Longhi}(2019)}]{PhysRevB.99.155150}%
  \BibitemOpen
  \bibfield  {author} {\bibinfo {author} {\bibfnamefont {S.}~\bibnamefont
  {Longhi}},\ }\href {\doibase 10.1103/PhysRevB.99.155150} {\bibfield
  {journal} {\bibinfo  {journal} {Phys. Rev. B}\ }\textbf {\bibinfo {volume}
  {99}},\ \bibinfo {pages} {155150} (\bibinfo {year} {2019})}\BibitemShut
  {NoStop}%
\bibitem [{\citenamefont {D'Angelis}\ \emph {et~al.}(2020)\citenamefont
  {D'Angelis}, \citenamefont {Pinheiro}, \citenamefont {Gu\'ery-Odelin},
  \citenamefont {Longhi},\ and\ \citenamefont
  {Impens}}]{PhysRevResearch.2.033475}%
  \BibitemOpen
  \bibfield  {author} {\bibinfo {author} {\bibfnamefont {F.~M.}\ \bibnamefont
  {D'Angelis}}, \bibinfo {author} {\bibfnamefont {F.~A.}\ \bibnamefont
  {Pinheiro}}, \bibinfo {author} {\bibfnamefont {D.}~\bibnamefont
  {Gu\'ery-Odelin}}, \bibinfo {author} {\bibfnamefont {S.}~\bibnamefont
  {Longhi}}, \ and\ \bibinfo {author} {\bibfnamefont {F.~m.~c.}\ \bibnamefont
  {Impens}},\ }\href {\doibase 10.1103/PhysRevResearch.2.033475} {\bibfield
  {journal} {\bibinfo  {journal} {Phys. Rev. Res.}\ }\textbf {\bibinfo {volume}
  {2}},\ \bibinfo {pages} {033475} (\bibinfo {year} {2020})}\BibitemShut
  {NoStop}%
\bibitem [{\citenamefont {Tian}\ \emph {et~al.}(2022)\citenamefont {Tian},
  \citenamefont {Zhang}, \citenamefont {Zhang}, \citenamefont {Wu},
  \citenamefont {Lin}, \citenamefont {Zhou}, \citenamefont {Duan},
  \citenamefont {Jiang},\ and\ \citenamefont {Du}}]{PhysRevLett.129.215901}%
  \BibitemOpen
  \bibfield  {author} {\bibinfo {author} {\bibfnamefont {T.}~\bibnamefont
  {Tian}}, \bibinfo {author} {\bibfnamefont {Y.}~\bibnamefont {Zhang}},
  \bibinfo {author} {\bibfnamefont {L.}~\bibnamefont {Zhang}}, \bibinfo
  {author} {\bibfnamefont {L.}~\bibnamefont {Wu}}, \bibinfo {author}
  {\bibfnamefont {S.}~\bibnamefont {Lin}}, \bibinfo {author} {\bibfnamefont
  {J.}~\bibnamefont {Zhou}}, \bibinfo {author} {\bibfnamefont {C.-K.}\
  \bibnamefont {Duan}}, \bibinfo {author} {\bibfnamefont {J.-H.}\ \bibnamefont
  {Jiang}}, \ and\ \bibinfo {author} {\bibfnamefont {J.}~\bibnamefont {Du}},\
  }\href {\doibase 10.1103/PhysRevLett.129.215901} {\bibfield  {journal}
  {\bibinfo  {journal} {Phys. Rev. Lett.}\ }\textbf {\bibinfo {volume} {129}},\
  \bibinfo {pages} {215901} (\bibinfo {year} {2022})}\BibitemShut {NoStop}%
\bibitem [{\citenamefont {Wang}\ \emph {et~al.}(2023)\citenamefont {Wang},
  \citenamefont {Zhao}, \citenamefont {Yang}, \citenamefont {Yan},\ and\
  \citenamefont {Zhou}}]{PhysRevA.107.053701}%
  \BibitemOpen
  \bibfield  {author} {\bibinfo {author} {\bibfnamefont {D.-W.}\ \bibnamefont
  {Wang}}, \bibinfo {author} {\bibfnamefont {C.}~\bibnamefont {Zhao}}, \bibinfo
  {author} {\bibfnamefont {J.}~\bibnamefont {Yang}}, \bibinfo {author}
  {\bibfnamefont {Y.-T.}\ \bibnamefont {Yan}}, \ and\ \bibinfo {author}
  {\bibfnamefont {L.}~\bibnamefont {Zhou}},\ }\href {\doibase
  10.1103/PhysRevA.107.053701} {\bibfield  {journal} {\bibinfo  {journal}
  {Phys. Rev. A}\ }\textbf {\bibinfo {volume} {107}},\ \bibinfo {pages}
  {053701} (\bibinfo {year} {2023})}\BibitemShut {NoStop}%
\bibitem [{\citenamefont {Wang}\ \emph {et~al.}(2022)\citenamefont {Wang},
  \citenamefont {Li}, \citenamefont {Gong},\ and\ \citenamefont
  {X.~Liu}}]{PhysRevA.106.052411}%
  \BibitemOpen
  \bibfield  {author} {\bibinfo {author} {\bibfnamefont {C.}~\bibnamefont
  {Wang}}, \bibinfo {author} {\bibfnamefont {L.}~\bibnamefont {Li}}, \bibinfo
  {author} {\bibfnamefont {J.}~\bibnamefont {Gong}}, \ and\ \bibinfo {author}
  {\bibfnamefont {Y.}~\bibnamefont {X.~Liu}},\ }\href {\doibase
  10.1103/PhysRevA.106.052411} {\bibfield  {journal} {\bibinfo  {journal}
  {Phys. Rev. A}\ }\textbf {\bibinfo {volume} {106}},\ \bibinfo {pages}
  {052411} (\bibinfo {year} {2022})}\BibitemShut {NoStop}%
\bibitem [{\citenamefont {Qi}\ \emph {et~al.}(2020)\citenamefont {Qi},
  \citenamefont {Wang}, \citenamefont {Liu}, \citenamefont {Zhang},\ and\
  \citenamefont {Wang}}]{PhysRevA.102.022404}%
  \BibitemOpen
  \bibfield  {author} {\bibinfo {author} {\bibfnamefont {L.}~\bibnamefont
  {Qi}}, \bibinfo {author} {\bibfnamefont {G.-L.}\ \bibnamefont {Wang}},
  \bibinfo {author} {\bibfnamefont {S.}~\bibnamefont {Liu}}, \bibinfo {author}
  {\bibfnamefont {S.}~\bibnamefont {Zhang}}, \ and\ \bibinfo {author}
  {\bibfnamefont {H.-F.}\ \bibnamefont {Wang}},\ }\href {\doibase
  10.1103/PhysRevA.102.022404} {\bibfield  {journal} {\bibinfo  {journal}
  {Phys. Rev. A}\ }\textbf {\bibinfo {volume} {102}},\ \bibinfo {pages}
  {022404} (\bibinfo {year} {2020})}\BibitemShut {NoStop}%
\bibitem [{\citenamefont {Cao}\ \emph {et~al.}(2021)\citenamefont {Cao},
  \citenamefont {Cui}, \citenamefont {Yi},\ and\ \citenamefont
  {Wang}}]{PhysRevA.103.023504}%
  \BibitemOpen
  \bibfield  {author} {\bibinfo {author} {\bibfnamefont {J.}~\bibnamefont
  {Cao}}, \bibinfo {author} {\bibfnamefont {W.-X.}\ \bibnamefont {Cui}},
  \bibinfo {author} {\bibfnamefont {X.~X.}\ \bibnamefont {Yi}}, \ and\ \bibinfo
  {author} {\bibfnamefont {H.-F.}\ \bibnamefont {Wang}},\ }\href {\doibase
  10.1103/PhysRevA.103.023504} {\bibfield  {journal} {\bibinfo  {journal}
  {Phys. Rev. A}\ }\textbf {\bibinfo {volume} {103}},\ \bibinfo {pages}
  {023504} (\bibinfo {year} {2021})}\BibitemShut {NoStop}%
\bibitem [{\citenamefont {Zheng}\ \emph {et~al.}(2022)\citenamefont {Zheng},
  \citenamefont {Yi},\ and\ \citenamefont {Wang}}]{PhysRevApplied.18.054037}%
  \BibitemOpen
  \bibfield  {author} {\bibinfo {author} {\bibfnamefont {L.-N.}\ \bibnamefont
  {Zheng}}, \bibinfo {author} {\bibfnamefont {X.}~\bibnamefont {Yi}}, \ and\
  \bibinfo {author} {\bibfnamefont {H.-F.}\ \bibnamefont {Wang}},\ }\href
  {\doibase 10.1103/PhysRevApplied.18.054037} {\bibfield  {journal} {\bibinfo
  {journal} {Phys. Rev. Appl.}\ }\textbf {\bibinfo {volume} {18}},\ \bibinfo
  {pages} {054037} (\bibinfo {year} {2022})}\BibitemShut {NoStop}%
\bibitem [{\citenamefont {Qi}\ \emph {et~al.}(2023)\citenamefont {Qi},
  \citenamefont {Han}, \citenamefont {Liu}, \citenamefont {Wang},\ and\
  \citenamefont {He}}]{PhysRevA.107.062214}%
  \BibitemOpen
  \bibfield  {author} {\bibinfo {author} {\bibfnamefont {L.}~\bibnamefont
  {Qi}}, \bibinfo {author} {\bibfnamefont {N.}~\bibnamefont {Han}}, \bibinfo
  {author} {\bibfnamefont {S.}~\bibnamefont {Liu}}, \bibinfo {author}
  {\bibfnamefont {H.-F.}\ \bibnamefont {Wang}}, \ and\ \bibinfo {author}
  {\bibfnamefont {A.-L.}\ \bibnamefont {He}},\ }\href {\doibase
  10.1103/PhysRevA.107.062214} {\bibfield  {journal} {\bibinfo  {journal}
  {Phys. Rev. A}\ }\textbf {\bibinfo {volume} {107}},\ \bibinfo {pages}
  {062214} (\bibinfo {year} {2023})}\BibitemShut {NoStop}%
\bibitem [{\citenamefont {Qi}\ \emph {et~al.}(2021{\natexlab{a}})\citenamefont
  {Qi}, \citenamefont {Yan}, \citenamefont {Xing}, \citenamefont {Zhao},
  \citenamefont {Liu}, \citenamefont {Cui}, \citenamefont {Han}, \citenamefont
  {Zhang},\ and\ \citenamefont {Wang}}]{PhysRevResearch.3.023037}%
  \BibitemOpen
  \bibfield  {author} {\bibinfo {author} {\bibfnamefont {L.}~\bibnamefont
  {Qi}}, \bibinfo {author} {\bibfnamefont {Y.}~\bibnamefont {Yan}}, \bibinfo
  {author} {\bibfnamefont {Y.}~\bibnamefont {Xing}}, \bibinfo {author}
  {\bibfnamefont {X.-D.}\ \bibnamefont {Zhao}}, \bibinfo {author}
  {\bibfnamefont {S.}~\bibnamefont {Liu}}, \bibinfo {author} {\bibfnamefont
  {W.-X.}\ \bibnamefont {Cui}}, \bibinfo {author} {\bibfnamefont
  {X.}~\bibnamefont {Han}}, \bibinfo {author} {\bibfnamefont {S.}~\bibnamefont
  {Zhang}}, \ and\ \bibinfo {author} {\bibfnamefont {H.-F.}\ \bibnamefont
  {Wang}},\ }\href {\doibase 10.1103/PhysRevResearch.3.023037} {\bibfield
  {journal} {\bibinfo  {journal} {Phys. Rev. Res.}\ }\textbf {\bibinfo {volume}
  {3}},\ \bibinfo {pages} {023037} (\bibinfo {year}
  {2021}{\natexlab{a}})}\BibitemShut {NoStop}%
\bibitem [{\citenamefont {Qi}\ \emph {et~al.}(2021{\natexlab{b}})\citenamefont
  {Qi}, \citenamefont {Xing}, \citenamefont {Zhao}, \citenamefont {Liu},
  \citenamefont {Zhang}, \citenamefont {Hu},\ and\ \citenamefont
  {Wang}}]{PhysRevB.103.085129}%
  \BibitemOpen
  \bibfield  {author} {\bibinfo {author} {\bibfnamefont {L.}~\bibnamefont
  {Qi}}, \bibinfo {author} {\bibfnamefont {Y.}~\bibnamefont {Xing}}, \bibinfo
  {author} {\bibfnamefont {X.-D.}\ \bibnamefont {Zhao}}, \bibinfo {author}
  {\bibfnamefont {S.}~\bibnamefont {Liu}}, \bibinfo {author} {\bibfnamefont
  {S.}~\bibnamefont {Zhang}}, \bibinfo {author} {\bibfnamefont
  {S.}~\bibnamefont {Hu}}, \ and\ \bibinfo {author} {\bibfnamefont {H.-F.}\
  \bibnamefont {Wang}},\ }\href {\doibase 10.1103/PhysRevB.103.085129}
  {\bibfield  {journal} {\bibinfo  {journal} {Phys. Rev. B}\ }\textbf {\bibinfo
  {volume} {103}},\ \bibinfo {pages} {085129} (\bibinfo {year}
  {2021}{\natexlab{b}})}\BibitemShut {NoStop}%
\bibitem [{\citenamefont {Ezawa}(2024)}]{PhysRevB.109.205421}%
  \BibitemOpen
  \bibfield  {author} {\bibinfo {author} {\bibfnamefont {M.}~\bibnamefont
  {Ezawa}},\ }\href {\doibase 10.1103/PhysRevB.109.205421} {\bibfield
  {journal} {\bibinfo  {journal} {Phys. Rev. B}\ }\textbf {\bibinfo {volume}
  {109}},\ \bibinfo {pages} {205421} (\bibinfo {year} {2024})}\BibitemShut
  {NoStop}%
\bibitem [{\citenamefont {Nie}\ \emph {et~al.}(2020)\citenamefont {Nie},
  \citenamefont {Peng}, \citenamefont {Nori},\ and\ \citenamefont
  {X.~Liu}}]{PhysRevLett.124.023603}%
  \BibitemOpen
  \bibfield  {author} {\bibinfo {author} {\bibfnamefont {W.}~\bibnamefont
  {Nie}}, \bibinfo {author} {\bibfnamefont {Z.~H.}\ \bibnamefont {Peng}},
  \bibinfo {author} {\bibfnamefont {F.}~\bibnamefont {Nori}}, \ and\ \bibinfo
  {author} {\bibfnamefont {Y.}~\bibnamefont {X.~Liu}},\ }\href {\doibase
  10.1103/PhysRevLett.124.023603} {\bibfield  {journal} {\bibinfo  {journal}
  {Phys. Rev. Lett.}\ }\textbf {\bibinfo {volume} {124}},\ \bibinfo {pages}
  {023603} (\bibinfo {year} {2020})}\BibitemShut {NoStop}%
\bibitem [{\citenamefont {Nie}\ and\ \citenamefont
  {X.~Liu}(2020)}]{PhysRevResearch.2.012076}%
  \BibitemOpen
  \bibfield  {author} {\bibinfo {author} {\bibfnamefont {W.}~\bibnamefont
  {Nie}}\ and\ \bibinfo {author} {\bibfnamefont {Y.}~\bibnamefont {X.~Liu}},\
  }\href {\doibase 10.1103/PhysRevResearch.2.012076} {\bibfield  {journal}
  {\bibinfo  {journal} {Phys. Rev. Res.}\ }\textbf {\bibinfo {volume} {2}},\
  \bibinfo {pages} {012076} (\bibinfo {year} {2020})}\BibitemShut {NoStop}%
\bibitem [{\citenamefont {Dlaska}\ \emph {et~al.}(2017)\citenamefont {Dlaska},
  \citenamefont {Vermersch},\ and\ \citenamefont {Zoller}}]{Dlaska_2017}%
  \BibitemOpen
  \bibfield  {author} {\bibinfo {author} {\bibfnamefont {C.}~\bibnamefont
  {Dlaska}}, \bibinfo {author} {\bibfnamefont {B.}~\bibnamefont {Vermersch}}, \
  and\ \bibinfo {author} {\bibfnamefont {P.}~\bibnamefont {Zoller}},\ }\href
  {\doibase 10.1088/2058-9565/2/1/015001} {\bibfield  {journal} {\bibinfo
  {journal} {Quantum Sci. Technol}\ }\textbf {\bibinfo {volume} {2}},\ \bibinfo
  {pages} {015001} (\bibinfo {year} {2017})}\BibitemShut {NoStop}%
\bibitem [{\citenamefont {Yao}\ \emph {et~al.}(2013)\citenamefont {Yao},
  \citenamefont {Laumann}, \citenamefont {Gorshkov}, \citenamefont {Weimer},
  \citenamefont {Jiang}, \citenamefont {Cirac}, \citenamefont {Zoller},\ and\
  \citenamefont {Lukin}}]{Yao2013}%
  \BibitemOpen
  \bibfield  {author} {\bibinfo {author} {\bibfnamefont {N.~Y.}\ \bibnamefont
  {Yao}}, \bibinfo {author} {\bibfnamefont {C.~R.}\ \bibnamefont {Laumann}},
  \bibinfo {author} {\bibfnamefont {A.~V.}\ \bibnamefont {Gorshkov}}, \bibinfo
  {author} {\bibfnamefont {H.}~\bibnamefont {Weimer}}, \bibinfo {author}
  {\bibfnamefont {L.}~\bibnamefont {Jiang}}, \bibinfo {author} {\bibfnamefont
  {J.~I.}\ \bibnamefont {Cirac}}, \bibinfo {author} {\bibfnamefont
  {P.}~\bibnamefont {Zoller}}, \ and\ \bibinfo {author} {\bibfnamefont {M.~D.}\
  \bibnamefont {Lukin}},\ }\href {\doibase 10.1038/ncomms2531} {\bibfield
  {journal} {\bibinfo  {journal} {Nat.Commun.}\ }\textbf {\bibinfo {volume}
  {4}},\ \bibinfo {pages} {1585} (\bibinfo {year} {2013})}\BibitemShut
  {NoStop}%
\bibitem [{\citenamefont {Wei}(2022)}]{PhysRevA.106.033710}%
  \BibitemOpen
  \bibfield  {author} {\bibinfo {author} {\bibfnamefont {J.}~\bibnamefont
  {Wei}},\ }\href {\doibase 10.1103/PhysRevA.106.033710} {\bibfield  {journal}
  {\bibinfo  {journal} {Phys. Rev. A}\ }\textbf {\bibinfo {volume} {106}},\
  \bibinfo {pages} {033710} (\bibinfo {year} {2022})}\BibitemShut {NoStop}%
\bibitem [{\citenamefont {Kvande}\ \emph {et~al.}(2023)\citenamefont {Kvande},
  \citenamefont {Hill},\ and\ \citenamefont {Blume}}]{PhysRevA.108.023703}%
  \BibitemOpen
  \bibfield  {author} {\bibinfo {author} {\bibfnamefont {C.~I.}\ \bibnamefont
  {Kvande}}, \bibinfo {author} {\bibfnamefont {D.~B.}\ \bibnamefont {Hill}}, \
  and\ \bibinfo {author} {\bibfnamefont {D.}~\bibnamefont {Blume}},\ }\href
  {\doibase 10.1103/PhysRevA.108.023703} {\bibfield  {journal} {\bibinfo
  {journal} {Phys. Rev. A}\ }\textbf {\bibinfo {volume} {108}},\ \bibinfo
  {pages} {023703} (\bibinfo {year} {2023})}\BibitemShut {NoStop}%
\bibitem [{\citenamefont {Wang}\ \emph {et~al.}(2024)\citenamefont {Wang},
  \citenamefont {Zhao}, \citenamefont {Yang}, \citenamefont {Yan},\ and\
  \citenamefont {Zhou}}]{PhysRevA.109.033708}%
  \BibitemOpen
  \bibfield  {author} {\bibinfo {author} {\bibfnamefont {D.-W.}\ \bibnamefont
  {Wang}}, \bibinfo {author} {\bibfnamefont {C.}~\bibnamefont {Zhao}}, \bibinfo
  {author} {\bibfnamefont {J.}~\bibnamefont {Yang}}, \bibinfo {author}
  {\bibfnamefont {Y.-T.}\ \bibnamefont {Yan}}, \ and\ \bibinfo {author}
  {\bibfnamefont {L.}~\bibnamefont {Zhou}},\ }\href {\doibase
  10.1103/PhysRevA.109.033708} {\bibfield  {journal} {\bibinfo  {journal}
  {Phys. Rev. A}\ }\textbf {\bibinfo {volume} {109}},\ \bibinfo {pages}
  {033708} (\bibinfo {year} {2024})}\BibitemShut {NoStop}%
\bibitem [{\citenamefont {Andersson}\ \emph {et~al.}(2019)\citenamefont
  {Andersson}, \citenamefont {Suri}, \citenamefont {Guo}, \citenamefont
  {Aref},\ and\ \citenamefont {Delsing}}]{Andersson2019}%
  \BibitemOpen
  \bibfield  {author} {\bibinfo {author} {\bibfnamefont {G.}~\bibnamefont
  {Andersson}}, \bibinfo {author} {\bibfnamefont {B.}~\bibnamefont {Suri}},
  \bibinfo {author} {\bibfnamefont {L.}~\bibnamefont {Guo}}, \bibinfo {author}
  {\bibfnamefont {T.}~\bibnamefont {Aref}}, \ and\ \bibinfo {author}
  {\bibfnamefont {P.}~\bibnamefont {Delsing}},\ }\href {\doibase
  10.1038/s41567-019-0605-6} {\bibfield  {journal} {\bibinfo  {journal} {Nat.
  Phys}\ }\textbf {\bibinfo {volume} {15}},\ \bibinfo {pages} {1123} (\bibinfo
  {year} {2019})}\BibitemShut {NoStop}%
\bibitem [{\citenamefont {Kannan}\ \emph {et~al.}(2020)\citenamefont {Kannan},
  \citenamefont {Ruckriegel}, \citenamefont {Campbell}, \citenamefont {{Frisk
  Kockum}}, \citenamefont {Braum{\"{u}}ller}, \citenamefont {Kim},
  \citenamefont {Kjaergaard}, \citenamefont {Krantz}, \citenamefont {Melville},
  \citenamefont {Niedzielski}, \citenamefont {Veps{\"{a}}l{\"{a}}inen},
  \citenamefont {Winik}, \citenamefont {Yoder}, \citenamefont {Nori},
  \citenamefont {Orlando}, \citenamefont {Gustavsson},\ and\ \citenamefont
  {Oliver}}]{Kannan2020}%
  \BibitemOpen
  \bibfield  {author} {\bibinfo {author} {\bibfnamefont {B.}~\bibnamefont
  {Kannan}}, \bibinfo {author} {\bibfnamefont {M.~J.}\ \bibnamefont
  {Ruckriegel}}, \bibinfo {author} {\bibfnamefont {D.~L.}\ \bibnamefont
  {Campbell}}, \bibinfo {author} {\bibfnamefont {A.}~\bibnamefont {{Frisk
  Kockum}}}, \bibinfo {author} {\bibfnamefont {J.}~\bibnamefont
  {Braum{\"{u}}ller}}, \bibinfo {author} {\bibfnamefont {D.~K.}\ \bibnamefont
  {Kim}}, \bibinfo {author} {\bibfnamefont {M.}~\bibnamefont {Kjaergaard}},
  \bibinfo {author} {\bibfnamefont {P.}~\bibnamefont {Krantz}}, \bibinfo
  {author} {\bibfnamefont {A.}~\bibnamefont {Melville}}, \bibinfo {author}
  {\bibfnamefont {B.~M.}\ \bibnamefont {Niedzielski}}, \bibinfo {author}
  {\bibfnamefont {A.}~\bibnamefont {Veps{\"{a}}l{\"{a}}inen}}, \bibinfo
  {author} {\bibfnamefont {R.}~\bibnamefont {Winik}}, \bibinfo {author}
  {\bibfnamefont {J.~L.}\ \bibnamefont {Yoder}}, \bibinfo {author}
  {\bibfnamefont {F.}~\bibnamefont {Nori}}, \bibinfo {author} {\bibfnamefont
  {T.~P.}\ \bibnamefont {Orlando}}, \bibinfo {author} {\bibfnamefont
  {S.}~\bibnamefont {Gustavsson}}, \ and\ \bibinfo {author} {\bibfnamefont
  {W.~D.}\ \bibnamefont {Oliver}},\ }\href {\doibase 10.1038/s41586-020-2529-9}
  {\bibfield  {journal} {\bibinfo  {journal} {Nature}\ }\textbf {\bibinfo
  {volume} {583}},\ \bibinfo {pages} {775} (\bibinfo {year}
  {2020})}\BibitemShut {NoStop}%
\bibitem [{\citenamefont {Joshi}\ \emph {et~al.}(2023)\citenamefont {Joshi},
  \citenamefont {Yang},\ and\ \citenamefont
  {Mirhosseini}}]{PhysRevX.13.021039}%
  \BibitemOpen
  \bibfield  {author} {\bibinfo {author} {\bibfnamefont {C.}~\bibnamefont
  {Joshi}}, \bibinfo {author} {\bibfnamefont {F.}~\bibnamefont {Yang}}, \ and\
  \bibinfo {author} {\bibfnamefont {M.}~\bibnamefont {Mirhosseini}},\ }\href
  {\doibase 10.1103/PhysRevX.13.021039} {\bibfield  {journal} {\bibinfo
  {journal} {Phys. Rev. X}\ }\textbf {\bibinfo {volume} {13}},\ \bibinfo
  {pages} {021039} (\bibinfo {year} {2023})}\BibitemShut {NoStop}%
\bibitem [{\citenamefont {Vadiraj}\ \emph {et~al.}(2021)\citenamefont
  {Vadiraj}, \citenamefont {Ask}, \citenamefont {McConkey}, \citenamefont
  {Nsanzineza}, \citenamefont {Chang}, \citenamefont {Kockum},\ and\
  \citenamefont {Wilson}}]{PhysRevA.103.023710}%
  \BibitemOpen
  \bibfield  {author} {\bibinfo {author} {\bibfnamefont {A.~M.}\ \bibnamefont
  {Vadiraj}}, \bibinfo {author} {\bibfnamefont {A.}~\bibnamefont {Ask}},
  \bibinfo {author} {\bibfnamefont {T.~G.}\ \bibnamefont {McConkey}}, \bibinfo
  {author} {\bibfnamefont {I.}~\bibnamefont {Nsanzineza}}, \bibinfo {author}
  {\bibfnamefont {C.~W.~S.}\ \bibnamefont {Chang}}, \bibinfo {author}
  {\bibfnamefont {A.~F.}\ \bibnamefont {Kockum}}, \ and\ \bibinfo {author}
  {\bibfnamefont {C.~M.}\ \bibnamefont {Wilson}},\ }\href {\doibase
  10.1103/PhysRevA.103.023710} {\bibfield  {journal} {\bibinfo  {journal}
  {Phys. Rev. A}\ }\textbf {\bibinfo {volume} {103}},\ \bibinfo {pages}
  {023710} (\bibinfo {year} {2021})}\BibitemShut {NoStop}%
\bibitem [{\citenamefont {Frisk~Kockum}\ \emph {et~al.}(2014)\citenamefont
  {Frisk~Kockum}, \citenamefont {Delsing},\ and\ \citenamefont
  {Johansson}}]{PhysRevA.90.013837}%
  \BibitemOpen
  \bibfield  {author} {\bibinfo {author} {\bibfnamefont {A.}~\bibnamefont
  {Frisk~Kockum}}, \bibinfo {author} {\bibfnamefont {P.}~\bibnamefont
  {Delsing}}, \ and\ \bibinfo {author} {\bibfnamefont {G.}~\bibnamefont
  {Johansson}},\ }\href {\doibase 10.1103/PhysRevA.90.013837} {\bibfield
  {journal} {\bibinfo  {journal} {Phys. Rev. A}\ }\textbf {\bibinfo {volume}
  {90}},\ \bibinfo {pages} {013837} (\bibinfo {year} {2014})}\BibitemShut
  {NoStop}%
\bibitem [{\citenamefont {Wen}\ \emph {et~al.}(2019)\citenamefont {Wen},
  \citenamefont {Lin}, \citenamefont {Kockum}, \citenamefont {Suri},
  \citenamefont {Ian}, \citenamefont {Chen}, \citenamefont {Mao}, \citenamefont
  {Chiu}, \citenamefont {Delsing}, \citenamefont {Nori}, \citenamefont {Lin},\
  and\ \citenamefont {Hoi}}]{PhysRevLett.123.233602}%
  \BibitemOpen
  \bibfield  {author} {\bibinfo {author} {\bibfnamefont {P.~Y.}\ \bibnamefont
  {Wen}}, \bibinfo {author} {\bibfnamefont {K.-T.}\ \bibnamefont {Lin}},
  \bibinfo {author} {\bibfnamefont {A.~F.}\ \bibnamefont {Kockum}}, \bibinfo
  {author} {\bibfnamefont {B.}~\bibnamefont {Suri}}, \bibinfo {author}
  {\bibfnamefont {H.}~\bibnamefont {Ian}}, \bibinfo {author} {\bibfnamefont
  {J.~C.}\ \bibnamefont {Chen}}, \bibinfo {author} {\bibfnamefont {S.~Y.}\
  \bibnamefont {Mao}}, \bibinfo {author} {\bibfnamefont {C.~C.}\ \bibnamefont
  {Chiu}}, \bibinfo {author} {\bibfnamefont {P.}~\bibnamefont {Delsing}},
  \bibinfo {author} {\bibfnamefont {F.}~\bibnamefont {Nori}}, \bibinfo {author}
  {\bibfnamefont {G.-D.}\ \bibnamefont {Lin}}, \ and\ \bibinfo {author}
  {\bibfnamefont {I.-C.}\ \bibnamefont {Hoi}},\ }\href {\doibase
  10.1103/PhysRevLett.123.233602} {\bibfield  {journal} {\bibinfo  {journal}
  {Phys. Rev. Lett.}\ }\textbf {\bibinfo {volume} {123}},\ \bibinfo {pages}
  {233602} (\bibinfo {year} {2019})}\BibitemShut {NoStop}%
\bibitem [{\citenamefont {Kockum}\ \emph {et~al.}(2018)\citenamefont {Kockum},
  \citenamefont {Johansson},\ and\ \citenamefont
  {Nori}}]{PhysRevLett.120.140404}%
  \BibitemOpen
  \bibfield  {author} {\bibinfo {author} {\bibfnamefont {A.~F.}\ \bibnamefont
  {Kockum}}, \bibinfo {author} {\bibfnamefont {G.}~\bibnamefont {Johansson}}, \
  and\ \bibinfo {author} {\bibfnamefont {F.}~\bibnamefont {Nori}},\ }\href
  {\doibase 10.1103/PhysRevLett.120.140404} {\bibfield  {journal} {\bibinfo
  {journal} {Phys. Rev. Lett.}\ }\textbf {\bibinfo {volume} {120}},\ \bibinfo
  {pages} {140404} (\bibinfo {year} {2018})}\BibitemShut {NoStop}%
\bibitem [{\citenamefont {Carollo}\ \emph {et~al.}(2020)\citenamefont
  {Carollo}, \citenamefont {Cilluffo},\ and\ \citenamefont
  {Ciccarello}}]{PhysRevResearch.2.043184}%
  \BibitemOpen
  \bibfield  {author} {\bibinfo {author} {\bibfnamefont {A.}~\bibnamefont
  {Carollo}}, \bibinfo {author} {\bibfnamefont {D.}~\bibnamefont {Cilluffo}}, \
  and\ \bibinfo {author} {\bibfnamefont {F.}~\bibnamefont {Ciccarello}},\
  }\href {\doibase 10.1103/PhysRevResearch.2.043184} {\bibfield  {journal}
  {\bibinfo  {journal} {Phys. Rev. Res.}\ }\textbf {\bibinfo {volume} {2}},\
  \bibinfo {pages} {043184} (\bibinfo {year} {2020})}\BibitemShut {NoStop}%
\bibitem [{\citenamefont {Du}\ \emph {et~al.}(2023)\citenamefont {Du},
  \citenamefont {Guo},\ and\ \citenamefont {Li}}]{PhysRevA.107.023705}%
  \BibitemOpen
  \bibfield  {author} {\bibinfo {author} {\bibfnamefont {L.}~\bibnamefont
  {Du}}, \bibinfo {author} {\bibfnamefont {L.}~\bibnamefont {Guo}}, \ and\
  \bibinfo {author} {\bibfnamefont {Y.}~\bibnamefont {Li}},\ }\href {\doibase
  10.1103/PhysRevA.107.023705} {\bibfield  {journal} {\bibinfo  {journal}
  {Phys. Rev. A}\ }\textbf {\bibinfo {volume} {107}},\ \bibinfo {pages}
  {023705} (\bibinfo {year} {2023})}\BibitemShut {NoStop}%
\bibitem [{\citenamefont {Soro}\ and\ \citenamefont
  {Kockum}(2022)}]{PhysRevA.105.023712}%
  \BibitemOpen
  \bibfield  {author} {\bibinfo {author} {\bibfnamefont {A.}~\bibnamefont
  {Soro}}\ and\ \bibinfo {author} {\bibfnamefont {A.~F.}\ \bibnamefont
  {Kockum}},\ }\href {\doibase 10.1103/PhysRevA.105.023712} {\bibfield
  {journal} {\bibinfo  {journal} {Phys. Rev. A}\ }\textbf {\bibinfo {volume}
  {105}},\ \bibinfo {pages} {023712} (\bibinfo {year} {2022})}\BibitemShut
  {NoStop}%
\bibitem [{\citenamefont {Wang}\ \emph {et~al.}(2021)\citenamefont {Wang},
  \citenamefont {Liu}, \citenamefont {Kockum}, \citenamefont {Li},\ and\
  \citenamefont {Nori}}]{PhysRevLett.126.043602}%
  \BibitemOpen
  \bibfield  {author} {\bibinfo {author} {\bibfnamefont {X.}~\bibnamefont
  {Wang}}, \bibinfo {author} {\bibfnamefont {T.}~\bibnamefont {Liu}}, \bibinfo
  {author} {\bibfnamefont {A.~F.}\ \bibnamefont {Kockum}}, \bibinfo {author}
  {\bibfnamefont {H.-R.}\ \bibnamefont {Li}}, \ and\ \bibinfo {author}
  {\bibfnamefont {F.}~\bibnamefont {Nori}},\ }\href {\doibase
  10.1103/PhysRevLett.126.043602} {\bibfield  {journal} {\bibinfo  {journal}
  {Phys. Rev. Lett.}\ }\textbf {\bibinfo {volume} {126}},\ \bibinfo {pages}
  {043602} (\bibinfo {year} {2021})}\BibitemShut {NoStop}%
\bibitem [{\citenamefont {Guo}\ \emph {et~al.}(2020{\natexlab{a}})\citenamefont
  {Guo}, \citenamefont {Wang}, \citenamefont {Purdy},\ and\ \citenamefont
  {Taylor}}]{PhysRevA.102.033706}%
  \BibitemOpen
  \bibfield  {author} {\bibinfo {author} {\bibfnamefont {S.}~\bibnamefont
  {Guo}}, \bibinfo {author} {\bibfnamefont {Y.}~\bibnamefont {Wang}}, \bibinfo
  {author} {\bibfnamefont {T.}~\bibnamefont {Purdy}}, \ and\ \bibinfo {author}
  {\bibfnamefont {J.}~\bibnamefont {Taylor}},\ }\href {\doibase
  10.1103/PhysRevA.102.033706} {\bibfield  {journal} {\bibinfo  {journal}
  {Phys. Rev. A}\ }\textbf {\bibinfo {volume} {102}},\ \bibinfo {pages}
  {033706} (\bibinfo {year} {2020}{\natexlab{a}})}\BibitemShut {NoStop}%
\bibitem [{\citenamefont {Guo}\ \emph {et~al.}(2020{\natexlab{b}})\citenamefont
  {Guo}, \citenamefont {Kockum}, \citenamefont {Marquardt},\ and\ \citenamefont
  {Johansson}}]{PhysRevResearch.2.043014}%
  \BibitemOpen
  \bibfield  {author} {\bibinfo {author} {\bibfnamefont {L.}~\bibnamefont
  {Guo}}, \bibinfo {author} {\bibfnamefont {A.~F.}\ \bibnamefont {Kockum}},
  \bibinfo {author} {\bibfnamefont {F.}~\bibnamefont {Marquardt}}, \ and\
  \bibinfo {author} {\bibfnamefont {G.}~\bibnamefont {Johansson}},\ }\href
  {\doibase 10.1103/PhysRevResearch.2.043014} {\bibfield  {journal} {\bibinfo
  {journal} {Phys. Rev. Res.}\ }\textbf {\bibinfo {volume} {2}},\ \bibinfo
  {pages} {043014} (\bibinfo {year} {2020}{\natexlab{b}})}\BibitemShut
  {NoStop}%
\bibitem [{\citenamefont {Roccati}\ and\ \citenamefont
  {Cilluffo}(2024)}]{PhysRevLett.133.063603}%
  \BibitemOpen
  \bibfield  {author} {\bibinfo {author} {\bibfnamefont {F.}~\bibnamefont
  {Roccati}}\ and\ \bibinfo {author} {\bibfnamefont {D.}~\bibnamefont
  {Cilluffo}},\ }\href {\doibase 10.1103/PhysRevLett.133.063603} {\bibfield
  {journal} {\bibinfo  {journal} {Phys. Rev. Lett.}\ }\textbf {\bibinfo
  {volume} {133}},\ \bibinfo {pages} {063603} (\bibinfo {year}
  {2024})}\BibitemShut {NoStop}%
\bibitem [{\citenamefont {Cheng}\ \emph {et~al.}(2022)\citenamefont {Cheng},
  \citenamefont {Wang},\ and\ \citenamefont {X.~Liu}}]{PhysRevA.106.033522}%
  \BibitemOpen
  \bibfield  {author} {\bibinfo {author} {\bibfnamefont {W.}~\bibnamefont
  {Cheng}}, \bibinfo {author} {\bibfnamefont {Z.}~\bibnamefont {Wang}}, \ and\
  \bibinfo {author} {\bibfnamefont {Y.}~\bibnamefont {X.~Liu}},\ }\href
  {\doibase 10.1103/PhysRevA.106.033522} {\bibfield  {journal} {\bibinfo
  {journal} {Phys. Rev. A}\ }\textbf {\bibinfo {volume} {106}},\ \bibinfo
  {pages} {033522} (\bibinfo {year} {2022})}\BibitemShut {NoStop}%
\bibitem [{\citenamefont {Chen}\ \emph {et~al.}(2022)\citenamefont {Chen},
  \citenamefont {Du}, \citenamefont {Guo}, \citenamefont {Wang}, \citenamefont
  {Zhang}, \citenamefont {Li},\ and\ \citenamefont {Wu}}]{Chen2022}%
  \BibitemOpen
  \bibfield  {author} {\bibinfo {author} {\bibfnamefont {Y.-T.}\ \bibnamefont
  {Chen}}, \bibinfo {author} {\bibfnamefont {L.}~\bibnamefont {Du}}, \bibinfo
  {author} {\bibfnamefont {L.}~\bibnamefont {Guo}}, \bibinfo {author}
  {\bibfnamefont {Z.}~\bibnamefont {Wang}}, \bibinfo {author} {\bibfnamefont
  {Y.}~\bibnamefont {Zhang}}, \bibinfo {author} {\bibfnamefont
  {Y.}~\bibnamefont {Li}}, \ and\ \bibinfo {author} {\bibfnamefont {J.-H.}\
  \bibnamefont {Wu}},\ }\href {\doibase 10.1038/s42005-022-00991-3} {\bibfield
  {journal} {\bibinfo  {journal} {Commun. Phys.}\ }\textbf {\bibinfo {volume}
  {5}},\ \bibinfo {pages} {215} (\bibinfo {year} {2022})}\BibitemShut {NoStop}%
\bibitem [{\citenamefont {Delplace}\ \emph {et~al.}(2011)\citenamefont
  {Delplace}, \citenamefont {Ullmo},\ and\ \citenamefont
  {Montambaux}}]{PhysRevB.84.195452}%
  \BibitemOpen
  \bibfield  {author} {\bibinfo {author} {\bibfnamefont {P.}~\bibnamefont
  {Delplace}}, \bibinfo {author} {\bibfnamefont {D.}~\bibnamefont {Ullmo}}, \
  and\ \bibinfo {author} {\bibfnamefont {G.}~\bibnamefont {Montambaux}},\
  }\href {\doibase 10.1103/PhysRevB.84.195452} {\bibfield  {journal} {\bibinfo
  {journal} {Phys. Rev. B}\ }\textbf {\bibinfo {volume} {84}},\ \bibinfo
  {pages} {195452} (\bibinfo {year} {2011})}\BibitemShut {NoStop}%
\bibitem [{\citenamefont {Zak}(1989)}]{PhysRevLett.62.2747}%
  \BibitemOpen
  \bibfield  {author} {\bibinfo {author} {\bibfnamefont {J.}~\bibnamefont
  {Zak}},\ }\href {\doibase 10.1103/PhysRevLett.62.2747} {\bibfield  {journal}
  {\bibinfo  {journal} {Phys. Rev. Lett.}\ }\textbf {\bibinfo {volume} {62}},\
  \bibinfo {pages} {2747} (\bibinfo {year} {1989})}\BibitemShut {NoStop}%
\bibitem [{\citenamefont {Asb{\'{o}}th}\ \emph {et~al.}()\citenamefont
  {Asb{\'{o}}th}, \citenamefont {Oroszl{\'{a}}ny},\ and\ \citenamefont
  {P{\'{a}}lyi}}]{Asboth2015}%
  \BibitemOpen
  \bibfield  {author} {\bibinfo {author} {\bibfnamefont {J.~K.}\ \bibnamefont
  {Asb{\'{o}}th}}, \bibinfo {author} {\bibfnamefont {L.}~\bibnamefont
  {Oroszl{\'{a}}ny}}, \ and\ \bibinfo {author} {\bibfnamefont {A.}~\bibnamefont
  {P{\'{a}}lyi}},\ }\href {\doibase 10.1007/978-3-319-25607-8} {}\Eprint
  {http://arxiv.org/abs/1509.02295} {arXiv:1509.02295} \BibitemShut {NoStop}%
\bibitem [{\citenamefont {Bernier}\ \emph {et~al.}(2018)\citenamefont
  {Bernier}, \citenamefont {T\'oth}, \citenamefont {Feofanov},\ and\
  \citenamefont {Kippenberg}}]{PhysRevA.98.023841}%
  \BibitemOpen
  \bibfield  {author} {\bibinfo {author} {\bibfnamefont {N.~R.}\ \bibnamefont
  {Bernier}}, \bibinfo {author} {\bibfnamefont {L.~D.}\ \bibnamefont {T\'oth}},
  \bibinfo {author} {\bibfnamefont {A.~K.}\ \bibnamefont {Feofanov}}, \ and\
  \bibinfo {author} {\bibfnamefont {T.~J.}\ \bibnamefont {Kippenberg}},\ }\href
  {\doibase 10.1103/PhysRevA.98.023841} {\bibfield  {journal} {\bibinfo
  {journal} {Phys. Rev. A}\ }\textbf {\bibinfo {volume} {98}},\ \bibinfo
  {pages} {023841} (\bibinfo {year} {2018})}\BibitemShut {NoStop}%
\bibitem [{\citenamefont {Bai}\ \emph {et~al.}(2015)\citenamefont {Bai},
  \citenamefont {Harder}, \citenamefont {Chen}, \citenamefont {Fan},
  \citenamefont {Xiao},\ and\ \citenamefont {Hu}}]{PhysRevLett.114.227201}%
  \BibitemOpen
  \bibfield  {author} {\bibinfo {author} {\bibfnamefont {L.}~\bibnamefont
  {Bai}}, \bibinfo {author} {\bibfnamefont {M.}~\bibnamefont {Harder}},
  \bibinfo {author} {\bibfnamefont {Y.~P.}\ \bibnamefont {Chen}}, \bibinfo
  {author} {\bibfnamefont {X.}~\bibnamefont {Fan}}, \bibinfo {author}
  {\bibfnamefont {J.~Q.}\ \bibnamefont {Xiao}}, \ and\ \bibinfo {author}
  {\bibfnamefont {C.-M.}\ \bibnamefont {Hu}},\ }\href {\doibase
  10.1103/PhysRevLett.114.227201} {\bibfield  {journal} {\bibinfo  {journal}
  {Phys. Rev. Lett.}\ }\textbf {\bibinfo {volume} {114}},\ \bibinfo {pages}
  {227201} (\bibinfo {year} {2015})}\BibitemShut {NoStop}%
\bibitem [{\citenamefont {Gu}\ \emph {et~al.}(2017)\citenamefont {Gu},
  \citenamefont {Kockum}, \citenamefont {Miranowicz}, \citenamefont {X.~Liu},\
  and\ \citenamefont {Nori}}]{Gu2017}%
  \BibitemOpen
  \bibfield  {author} {\bibinfo {author} {\bibfnamefont {X.}~\bibnamefont
  {Gu}}, \bibinfo {author} {\bibfnamefont {A.~F.}\ \bibnamefont {Kockum}},
  \bibinfo {author} {\bibfnamefont {A.}~\bibnamefont {Miranowicz}}, \bibinfo
  {author} {\bibfnamefont {Y.}~\bibnamefont {X.~Liu}}, \ and\ \bibinfo {author}
  {\bibfnamefont {F.}~\bibnamefont {Nori}},\ }\href {\doibase
  10.1016/j.physrep.2017.10.002} {\bibfield  {journal} {\bibinfo  {journal}
  {Phys. Rep.}\ }\textbf {\bibinfo {volume} {718-719}},\ \bibinfo {pages} {1}
  (\bibinfo {year} {2017})}\BibitemShut {NoStop}%
\bibitem [{\citenamefont {Krantz}\ \emph {et~al.}(2019)\citenamefont {Krantz},
  \citenamefont {Kjaergaard}, \citenamefont {Yan}, \citenamefont {Orlando},
  \citenamefont {Gustavsson},\ and\ \citenamefont
  {Oliver}}]{10.1063/1.5089550}%
  \BibitemOpen
  \bibfield  {author} {\bibinfo {author} {\bibfnamefont {P.}~\bibnamefont
  {Krantz}}, \bibinfo {author} {\bibfnamefont {M.}~\bibnamefont {Kjaergaard}},
  \bibinfo {author} {\bibfnamefont {F.}~\bibnamefont {Yan}}, \bibinfo {author}
  {\bibfnamefont {T.~P.}\ \bibnamefont {Orlando}}, \bibinfo {author}
  {\bibfnamefont {S.}~\bibnamefont {Gustavsson}}, \ and\ \bibinfo {author}
  {\bibfnamefont {W.~D.}\ \bibnamefont {Oliver}},\ }\href {\doibase
  10.1063/1.5089550} {\bibfield  {journal} {\bibinfo  {journal} {Appl. Phys.
  Rev.}\ }\textbf {\bibinfo {volume} {6}},\ \bibinfo {pages} {021318} (\bibinfo
  {year} {2019})}\BibitemShut {NoStop}%
\bibitem [{\citenamefont {Blais}\ \emph {et~al.}(2021)\citenamefont {Blais},
  \citenamefont {Grimsmo}, \citenamefont {Girvin},\ and\ \citenamefont
  {Wallraff}}]{RevModPhys.93.025005}%
  \BibitemOpen
  \bibfield  {author} {\bibinfo {author} {\bibfnamefont {A.}~\bibnamefont
  {Blais}}, \bibinfo {author} {\bibfnamefont {A.~L.}\ \bibnamefont {Grimsmo}},
  \bibinfo {author} {\bibfnamefont {S.~M.}\ \bibnamefont {Girvin}}, \ and\
  \bibinfo {author} {\bibfnamefont {A.}~\bibnamefont {Wallraff}},\ }\href
  {\doibase 10.1103/RevModPhys.93.025005} {\bibfield  {journal} {\bibinfo
  {journal} {Rev. Mod. Phys.}\ }\textbf {\bibinfo {volume} {93}},\ \bibinfo
  {pages} {025005} (\bibinfo {year} {2021})}\BibitemShut {NoStop}%
\bibitem [{\citenamefont {Vion}\ \emph {et~al.}(2002)\citenamefont {Vion},
  \citenamefont {Aassime}, \citenamefont {Cottet}, \citenamefont {Joyez},
  \citenamefont {Pothier}, \citenamefont {Urbina}, \citenamefont {Esteve},\
  and\ \citenamefont {Devoret}}]{doi:10.1126/science.1069372}%
  \BibitemOpen
  \bibfield  {author} {\bibinfo {author} {\bibfnamefont {D.}~\bibnamefont
  {Vion}}, \bibinfo {author} {\bibfnamefont {A.}~\bibnamefont {Aassime}},
  \bibinfo {author} {\bibfnamefont {A.}~\bibnamefont {Cottet}}, \bibinfo
  {author} {\bibfnamefont {P.}~\bibnamefont {Joyez}}, \bibinfo {author}
  {\bibfnamefont {H.}~\bibnamefont {Pothier}}, \bibinfo {author} {\bibfnamefont
  {C.}~\bibnamefont {Urbina}}, \bibinfo {author} {\bibfnamefont
  {D.}~\bibnamefont {Esteve}}, \ and\ \bibinfo {author} {\bibfnamefont {M.~H.}\
  \bibnamefont {Devoret}},\ }\href
  {https://www.science.org/doi/abs/10.1126/science.1069372} {\bibfield
  {journal} {\bibinfo  {journal} {Science}\ }\textbf {\bibinfo {volume}
  {296}},\ \bibinfo {pages} {886} (\bibinfo {year} {2002})}\BibitemShut
  {NoStop}%
\bibitem [{\citenamefont {Buluta}\ and\ \citenamefont
  {Nori}(2009)}]{doi:10.1126/science.1177838}%
  \BibitemOpen
  \bibfield  {author} {\bibinfo {author} {\bibfnamefont {I.}~\bibnamefont
  {Buluta}}\ and\ \bibinfo {author} {\bibfnamefont {F.}~\bibnamefont {Nori}},\
  }\href {https://www.science.org/doi/abs/10.1126/science.1177838} {\bibfield
  {journal} {\bibinfo  {journal} {Science}\ }\textbf {\bibinfo {volume}
  {326}},\ \bibinfo {pages} {108} (\bibinfo {year} {2009})}\BibitemShut
  {NoStop}%
\bibitem [{\citenamefont {Fitzpatrick}\ \emph {et~al.}(2017)\citenamefont
  {Fitzpatrick}, \citenamefont {Sundaresan}, \citenamefont {Li}, \citenamefont
  {Koch},\ and\ \citenamefont {Houck}}]{PhysRevX.7.011016}%
  \BibitemOpen
  \bibfield  {author} {\bibinfo {author} {\bibfnamefont {M.}~\bibnamefont
  {Fitzpatrick}}, \bibinfo {author} {\bibfnamefont {N.~M.}\ \bibnamefont
  {Sundaresan}}, \bibinfo {author} {\bibfnamefont {A.~C.~Y.}\ \bibnamefont
  {Li}}, \bibinfo {author} {\bibfnamefont {J.}~\bibnamefont {Koch}}, \ and\
  \bibinfo {author} {\bibfnamefont {A.~A.}\ \bibnamefont {Houck}},\ }\href
  {\doibase 10.1103/PhysRevX.7.011016} {\bibfield  {journal} {\bibinfo
  {journal} {Phys. Rev. X}\ }\textbf {\bibinfo {volume} {7}},\ \bibinfo {pages}
  {011016} (\bibinfo {year} {2017})}\BibitemShut {NoStop}%
\bibitem [{\citenamefont {Noh}\ and\ \citenamefont
  {Angelakis}(2016)}]{Noh2017}%
  \BibitemOpen
  \bibfield  {author} {\bibinfo {author} {\bibfnamefont {C.}~\bibnamefont
  {Noh}}\ and\ \bibinfo {author} {\bibfnamefont {D.~G.}\ \bibnamefont
  {Angelakis}},\ }\href {\doibase 10.1088/0034-4885/80/1/016401} {\bibfield
  {journal} {\bibinfo  {journal} {Rep. Prog. Phys.}\ }\textbf {\bibinfo
  {volume} {80}},\ \bibinfo {pages} {016401} (\bibinfo {year}
  {2016})}\BibitemShut {NoStop}%
\bibitem [{\citenamefont {Daley}\ \emph {et~al.}(2022)\citenamefont {Daley},
  \citenamefont {Bloch}, \citenamefont {Kokail}, \citenamefont {Flannigan},
  \citenamefont {Pearson}, \citenamefont {Troyer},\ and\ \citenamefont
  {Zoller}}]{Daley2022}%
  \BibitemOpen
  \bibfield  {author} {\bibinfo {author} {\bibfnamefont {A.~J.}\ \bibnamefont
  {Daley}}, \bibinfo {author} {\bibfnamefont {I.}~\bibnamefont {Bloch}},
  \bibinfo {author} {\bibfnamefont {C.}~\bibnamefont {Kokail}}, \bibinfo
  {author} {\bibfnamefont {S.}~\bibnamefont {Flannigan}}, \bibinfo {author}
  {\bibfnamefont {N.}~\bibnamefont {Pearson}}, \bibinfo {author} {\bibfnamefont
  {M.}~\bibnamefont {Troyer}}, \ and\ \bibinfo {author} {\bibfnamefont
  {P.}~\bibnamefont {Zoller}},\ }\href {\doibase 10.1038/s41586-022-04940-6}
  {\bibfield  {journal} {\bibinfo  {journal} {Nature}\ }\textbf {\bibinfo
  {volume} {607}},\ \bibinfo {pages} {667} (\bibinfo {year}
  {2022})}\BibitemShut {NoStop}%
\bibitem [{\citenamefont {Blais}\ \emph {et~al.}(2007)\citenamefont {Blais},
  \citenamefont {Gambetta}, \citenamefont {Wallraff}, \citenamefont {Schuster},
  \citenamefont {Girvin}, \citenamefont {Devoret},\ and\ \citenamefont
  {Schoelkopf}}]{PhysRevA.75.032329}%
  \BibitemOpen
  \bibfield  {author} {\bibinfo {author} {\bibfnamefont {A.}~\bibnamefont
  {Blais}}, \bibinfo {author} {\bibfnamefont {J.}~\bibnamefont {Gambetta}},
  \bibinfo {author} {\bibfnamefont {A.}~\bibnamefont {Wallraff}}, \bibinfo
  {author} {\bibfnamefont {D.~I.}\ \bibnamefont {Schuster}}, \bibinfo {author}
  {\bibfnamefont {S.~M.}\ \bibnamefont {Girvin}}, \bibinfo {author}
  {\bibfnamefont {M.~H.}\ \bibnamefont {Devoret}}, \ and\ \bibinfo {author}
  {\bibfnamefont {R.~J.}\ \bibnamefont {Schoelkopf}},\ }\href {\doibase
  10.1103/PhysRevA.75.032329} {\bibfield  {journal} {\bibinfo  {journal} {Phys.
  Rev. A}\ }\textbf {\bibinfo {volume} {75}},\ \bibinfo {pages} {032329}
  (\bibinfo {year} {2007})}\BibitemShut {NoStop}%
\bibitem [{\citenamefont {Mirhosseini}\ \emph {et~al.}(2019)\citenamefont
  {Mirhosseini}, \citenamefont {Kim}, \citenamefont {Zhang}, \citenamefont
  {Sipahigil}, \citenamefont {Dieterle}, \citenamefont {Keller}, \citenamefont
  {Asenjo-Garcia}, \citenamefont {Chang},\ and\ \citenamefont
  {Painter}}]{Mirhosseini2019}%
  \BibitemOpen
  \bibfield  {author} {\bibinfo {author} {\bibfnamefont {M.}~\bibnamefont
  {Mirhosseini}}, \bibinfo {author} {\bibfnamefont {E.}~\bibnamefont {Kim}},
  \bibinfo {author} {\bibfnamefont {X.}~\bibnamefont {Zhang}}, \bibinfo
  {author} {\bibfnamefont {A.}~\bibnamefont {Sipahigil}}, \bibinfo {author}
  {\bibfnamefont {P.~B.}\ \bibnamefont {Dieterle}}, \bibinfo {author}
  {\bibfnamefont {A.~J.}\ \bibnamefont {Keller}}, \bibinfo {author}
  {\bibfnamefont {A.}~\bibnamefont {Asenjo-Garcia}}, \bibinfo {author}
  {\bibfnamefont {D.~E.}\ \bibnamefont {Chang}}, \ and\ \bibinfo {author}
  {\bibfnamefont {O.}~\bibnamefont {Painter}},\ }\href {\doibase
  10.1038/s41586-019-1196-1} {\bibfield  {journal} {\bibinfo  {journal}
  {Nature}\ }\textbf {\bibinfo {volume} {569}},\ \bibinfo {pages} {692}
  (\bibinfo {year} {2019})}\BibitemShut {NoStop}%
\bibitem [{\citenamefont {Tao}\ \emph {et~al.}()\citenamefont {Tao},
  \citenamefont {Huang}, \citenamefont {Niu}, \citenamefont {Zhang},
  \citenamefont {Ke}, \citenamefont {Gu}, \citenamefont {Lin}, \citenamefont
  {Qiu}, \citenamefont {Sun}, \citenamefont {Yang}, \citenamefont {Zhang},
  \citenamefont {Zhang}, \citenamefont {Zhao}, \citenamefont {Zhou},
  \citenamefont {Deng}, \citenamefont {Hu}, \citenamefont {Hu}, \citenamefont
  {Li}, \citenamefont {Liu}, \citenamefont {Tan}, \citenamefont {Xu},
  \citenamefont {Yan}, \citenamefont {Chen}, \citenamefont {Lee}, \citenamefont
  {Zhong}, \citenamefont {Liu},\ and\ \citenamefont {Yu}}]{tao2023}%
  \BibitemOpen
  \bibfield  {author} {\bibinfo {author} {\bibfnamefont {Z.}~\bibnamefont
  {Tao}}, \bibinfo {author} {\bibfnamefont {W.}~\bibnamefont {Huang}}, \bibinfo
  {author} {\bibfnamefont {J.}~\bibnamefont {Niu}}, \bibinfo {author}
  {\bibfnamefont {L.}~\bibnamefont {Zhang}}, \bibinfo {author} {\bibfnamefont
  {Y.}~\bibnamefont {Ke}}, \bibinfo {author} {\bibfnamefont {X.}~\bibnamefont
  {Gu}}, \bibinfo {author} {\bibfnamefont {L.}~\bibnamefont {Lin}}, \bibinfo
  {author} {\bibfnamefont {J.}~\bibnamefont {Qiu}}, \bibinfo {author}
  {\bibfnamefont {X.}~\bibnamefont {Sun}}, \bibinfo {author} {\bibfnamefont
  {X.}~\bibnamefont {Yang}}, \bibinfo {author} {\bibfnamefont {J.}~\bibnamefont
  {Zhang}}, \bibinfo {author} {\bibfnamefont {J.}~\bibnamefont {Zhang}},
  \bibinfo {author} {\bibfnamefont {S.}~\bibnamefont {Zhao}}, \bibinfo {author}
  {\bibfnamefont {Y.}~\bibnamefont {Zhou}}, \bibinfo {author} {\bibfnamefont
  {X.}~\bibnamefont {Deng}}, \bibinfo {author} {\bibfnamefont {C.}~\bibnamefont
  {Hu}}, \bibinfo {author} {\bibfnamefont {L.}~\bibnamefont {Hu}}, \bibinfo
  {author} {\bibfnamefont {J.}~\bibnamefont {Li}}, \bibinfo {author}
  {\bibfnamefont {Y.}~\bibnamefont {Liu}}, \bibinfo {author} {\bibfnamefont
  {D.}~\bibnamefont {Tan}}, \bibinfo {author} {\bibfnamefont {Y.}~\bibnamefont
  {Xu}}, \bibinfo {author} {\bibfnamefont {T.}~\bibnamefont {Yan}}, \bibinfo
  {author} {\bibfnamefont {Y.}~\bibnamefont {Chen}}, \bibinfo {author}
  {\bibfnamefont {C.}~\bibnamefont {Lee}}, \bibinfo {author} {\bibfnamefont
  {Y.}~\bibnamefont {Zhong}}, \bibinfo {author} {\bibfnamefont
  {S.}~\bibnamefont {Liu}}, \ and\ \bibinfo {author} {\bibfnamefont
  {D.}~\bibnamefont {Yu}},\ }\href {https://arxiv.org/abs/2303.04582} {}\Eprint
  {http://arxiv.org/abs/2303.04582} {arXiv:2303.04582} \BibitemShut {NoStop}%
\bibitem [{\citenamefont {Gu}\ \emph {et~al.}()\citenamefont {Gu},
  \citenamefont {Chen},\ and\ \citenamefont
  {Liu}}]{gu2017topologicaledgestatespumping}%
  \BibitemOpen
  \bibfield  {author} {\bibinfo {author} {\bibfnamefont {X.}~\bibnamefont
  {Gu}}, \bibinfo {author} {\bibfnamefont {S.}~\bibnamefont {Chen}}, \ and\
  \bibinfo {author} {\bibfnamefont {Y.~X.}\ \bibnamefont {Liu}},\ }\href
  {https://arxiv.org/abs/1711.06829} {}\Eprint
  {http://arxiv.org/abs/1711.06829} {arXiv:1711.06829} \BibitemShut {NoStop}%
\bibitem [{\citenamefont {Kim}\ \emph {et~al.}(2021)\citenamefont {Kim},
  \citenamefont {Zhang}, \citenamefont {Ferreira}, \citenamefont {Banker},
  \citenamefont {Iverson}, \citenamefont {Sipahigil}, \citenamefont {Bello},
  \citenamefont {Gonz\'alez-Tudela}, \citenamefont {Mirhosseini},\ and\
  \citenamefont {Painter}}]{PhysRevX.11.011015}%
  \BibitemOpen
  \bibfield  {author} {\bibinfo {author} {\bibfnamefont {E.}~\bibnamefont
  {Kim}}, \bibinfo {author} {\bibfnamefont {X.}~\bibnamefont {Zhang}}, \bibinfo
  {author} {\bibfnamefont {V.~S.}\ \bibnamefont {Ferreira}}, \bibinfo {author}
  {\bibfnamefont {J.}~\bibnamefont {Banker}}, \bibinfo {author} {\bibfnamefont
  {J.~K.}\ \bibnamefont {Iverson}}, \bibinfo {author} {\bibfnamefont
  {A.}~\bibnamefont {Sipahigil}}, \bibinfo {author} {\bibfnamefont
  {M.}~\bibnamefont {Bello}}, \bibinfo {author} {\bibfnamefont
  {A.}~\bibnamefont {Gonz\'alez-Tudela}}, \bibinfo {author} {\bibfnamefont
  {M.}~\bibnamefont {Mirhosseini}}, \ and\ \bibinfo {author} {\bibfnamefont
  {O.}~\bibnamefont {Painter}},\ }\href {\doibase 10.1103/PhysRevX.11.011015}
  {\bibfield  {journal} {\bibinfo  {journal} {Phys. Rev. X}\ }\textbf {\bibinfo
  {volume} {11}},\ \bibinfo {pages} {011015} (\bibinfo {year}
  {2021})}\BibitemShut {NoStop}%
\bibitem [{\citenamefont {Bello}\ \emph {et~al.}(2019)\citenamefont {Bello},
  \citenamefont {Platero}, \citenamefont {Cirac},\ and\ \citenamefont
  {Gonz{\'{a}}lez-Tudela}}]{Bello2019}%
  \BibitemOpen
  \bibfield  {author} {\bibinfo {author} {\bibfnamefont {M.}~\bibnamefont
  {Bello}}, \bibinfo {author} {\bibfnamefont {G.}~\bibnamefont {Platero}},
  \bibinfo {author} {\bibfnamefont {J.~I.}\ \bibnamefont {Cirac}}, \ and\
  \bibinfo {author} {\bibfnamefont {A.}~\bibnamefont {Gonz{\'{a}}lez-Tudela}},\
  }\href {\doibase 10.1126/sciadv.aaw0297} {\bibfield  {journal} {\bibinfo
  {journal} {Sci. Adv.}\ }\textbf {\bibinfo {volume} {5}},\ \bibinfo {pages}
  {1} (\bibinfo {year} {2019})},\ \Eprint {http://arxiv.org/abs/1811.04390}
  {1811.04390} \BibitemShut {NoStop}%
\bibitem [{\citenamefont {Leonforte}\ \emph {et~al.}(2021)\citenamefont
  {Leonforte}, \citenamefont {Carollo},\ and\ \citenamefont
  {Ciccarello}}]{PhysRevLett.126.063601}%
  \BibitemOpen
  \bibfield  {author} {\bibinfo {author} {\bibfnamefont {L.}~\bibnamefont
  {Leonforte}}, \bibinfo {author} {\bibfnamefont {A.}~\bibnamefont {Carollo}},
  \ and\ \bibinfo {author} {\bibfnamefont {F.}~\bibnamefont {Ciccarello}},\
  }\href {\doibase 10.1103/PhysRevLett.126.063601} {\bibfield  {journal}
  {\bibinfo  {journal} {Phys. Rev. Lett.}\ }\textbf {\bibinfo {volume} {126}},\
  \bibinfo {pages} {063601} (\bibinfo {year} {2021})}\BibitemShut {NoStop}%
\bibitem [{\citenamefont {Houck}\ \emph {et~al.}(2012)\citenamefont {Houck},
  \citenamefont {T{\"{u}}reci},\ and\ \citenamefont {Koch}}]{Houck2012}%
  \BibitemOpen
  \bibfield  {author} {\bibinfo {author} {\bibfnamefont {A.~A.}\ \bibnamefont
  {Houck}}, \bibinfo {author} {\bibfnamefont {H.~E.}\ \bibnamefont
  {T{\"{u}}reci}}, \ and\ \bibinfo {author} {\bibfnamefont {J.}~\bibnamefont
  {Koch}},\ }\href {\doibase 10.1038/nphys2251} {\bibfield  {journal} {\bibinfo
   {journal} {Nat. Phys.}\ }\textbf {\bibinfo {volume} {8}},\ \bibinfo {pages}
  {292} (\bibinfo {year} {2012})}\BibitemShut {NoStop}%
\bibitem [{\citenamefont {Schmidt}\ and\ \citenamefont
  {Koch}(2013)}]{Schmidt2013}%
  \BibitemOpen
  \bibfield  {author} {\bibinfo {author} {\bibfnamefont {S.}~\bibnamefont
  {Schmidt}}\ and\ \bibinfo {author} {\bibfnamefont {J.}~\bibnamefont {Koch}},\
  }\href {\doibase 10.1002/andp.201200261} {\bibfield  {journal} {\bibinfo
  {journal} {Ann. Phys.}\ }\textbf {\bibinfo {volume} {525}},\ \bibinfo {pages}
  {395} (\bibinfo {year} {2013})}\BibitemShut {NoStop}%
\bibitem [{\citenamefont {Somoroff}\ \emph {et~al.}(2023)\citenamefont
  {Somoroff}, \citenamefont {Ficheux}, \citenamefont {Mencia}, \citenamefont
  {Xiong}, \citenamefont {Kuzmin},\ and\ \citenamefont
  {Manucharyan}}]{PhysRevLett.130.267001}%
  \BibitemOpen
  \bibfield  {author} {\bibinfo {author} {\bibfnamefont {A.}~\bibnamefont
  {Somoroff}}, \bibinfo {author} {\bibfnamefont {Q.}~\bibnamefont {Ficheux}},
  \bibinfo {author} {\bibfnamefont {R.~A.}\ \bibnamefont {Mencia}}, \bibinfo
  {author} {\bibfnamefont {H.}~\bibnamefont {Xiong}}, \bibinfo {author}
  {\bibfnamefont {R.}~\bibnamefont {Kuzmin}}, \ and\ \bibinfo {author}
  {\bibfnamefont {V.~E.}\ \bibnamefont {Manucharyan}},\ }\href {\doibase
  10.1103/PhysRevLett.130.267001} {\bibfield  {journal} {\bibinfo  {journal}
  {Phys. Rev. Lett.}\ }\textbf {\bibinfo {volume} {130}},\ \bibinfo {pages}
  {267001} (\bibinfo {year} {2023})}\BibitemShut {NoStop}%
\end{thebibliography}%
\end{document}